\documentclass[english,prd,twocolumn,superscriptaddress,floatfix,nofootinbib,preprintnumbers]{revtex4}
\usepackage{graphicx}
\usepackage{amsmath,amsthm,amssymb}
\usepackage{bm}

\usepackage{amsfonts}
\usepackage{dcolumn}
\usepackage{hyperref}

\def\be{\begin{equation}}
\def\ee{\end{equation}}
\def\ba{\begin{eqnarray}}
\def\ea{\end{eqnarray}}
\def\bs{\begin{subequations}}
\def\es{\end{subequations}}

\usepackage{color}

\makeatother

\usepackage{babel}
\makeatother

\begin{document}

\title{Constraints on $f(R)$ theory and Galileons from
\\ the latest data of galaxy redshift surveys}

\author{Hiroyuki Okada}
\affiliation{Department of Astronomy, Kyoto University, 
Kitashirakawa-Oiwake-cho, Sakyo-ku, Kyoto 606-8502, Japan}

\author{Tomonori Totani}
\affiliation{Department of Astronomy, Kyoto University, 
Kitashirakawa-Oiwake-cho, Sakyo-ku, Kyoto 606-8502, Japan}
\affiliation{Department of Astronomy, The University of Tokyo,
Hongo, Bunkyo-ku, Tokyo 113-0033, Japan }

\author{Shinji Tsujikawa}
\affiliation{Department of Physics, Faculty of Science, 
Tokyo University of Science,
1-3, Kagurazaka, Shinjuku-ku, Tokyo 162-8601, Japan}

\begin{abstract}
  The growth rate of matter density perturbations has been
  measured from redshift-space distortion (RSD) in the galaxy power
  spectrum.  We constrain the model parameter space for representative
  modified gravity models to explain the dark energy problem by using
  the recent data of $f_m(z) \sigma_8(z)$ at the redshifts $z = $ 0.06--0.8 
  measured by WiggleZ, SDSS LRG, BOSS, and 6dFGRS.  
  We first test the Hu-Sawicki's
  $f(R)$ dark energy model, and find that only the parameter region
  close to the standard $\Lambda$ Cold Dark Matter ($\Lambda$CDM)
  model is allowed ($\lambda > $ 12 and 5 for $n = $ 1.5 and 2,
  respectively, at 95\% CL).  We then investigate the covariant
  Galileon model with a de Sitter attractor and show that 
  the parameter space consistent with the background expansion history 
  is excluded by the RSD data at more than $8 \sigma$ because of the 
  too large growth rate predicted by the theory.  
  Finally, we consider the extended Galileon scenario, and
  we find that, in contrast to the covariant Galileon, there is a
  model parameter space for a tracker solution that is consistent with
  the RSD data within a $2\sigma$ level.
\end{abstract}

\date{\today}


\maketitle

\section{Introduction}

The observational support for the existence of dark energy
\cite{SN98,CMB03,LSS04} has motivated the idea that the gravitational
law may be modified from General Relativity (GR) at large distances to
realize the late-time cosmic acceleration 
(see Refs.~\cite{Sotiriou,DT10,fRde,Clifton} for reviews). 
In this vein many modified gravity
models of dark energy have been proposed---including those based on
$f(R)$ gravity \cite{fR}, scalar-tensor theories \cite{stensor}, the
Dvali-Gabadadze-Porrati (DGP) braneworld scenario \cite{DGP}, 
Galileons \cite{Nicolis,Deffayet}, and so on.

Measuring the growth rate of large scale structures in the Universe
gives a strong test for the modified gravity scenario, because
modified gravity models generally predict different growth rates from
that in the standard $\Lambda$ Cold Dark Matter ($\Lambda$CDM) model.
In fact the equations of linear matter density perturbations have been
derived for a number of modified gravity models---including $f(R)$
gravity \cite{fRper,Tsujikawa07}, the DGP model \cite{DGPper}, and
Galileons \cite{Kase}.  Recently De Felice {\it et al.} \cite{DKT}
derived the full perturbation equations as well as the effective
gravitational coupling to non-relativistic matter in the most general
scalar-tensor theories in 4 dimensions \cite{Horndeski,DGSZ,KYY11}
(which cover most of the single-field dark energy models proposed in
the literature).  

One of the methods to measure the cosmic growth rate is redshift-space
distortion (RSD) that appears in clustering pattern of galaxies in
galaxy redshift surveys because of radial peculiar velocities. 
RSD on large and linear scales reflects the velocity of inward
collapse motion of large scale structure, which is directly related
to the evolutionary speed of matter overdensity perturbations simply from 
the mass conservation \cite{Kaiser}.  Recent galaxy redshift
surveys have provided measurements of the growth rate $f_m(z)$ or
$f_m(z) \sigma_8(z)$ as a function of redshift up 
to $z \sim$ 1 \cite{Tegmark04,Percival04,Tegmark06,Yamamoto,Guzzo08,Blake,Samushia11,Reid12,Beutler12,Samushia12,Hudson,Jailson},
where $f_m = d\ln \delta_m/d\ln a$, $\delta_m$ is the fractional
over-density of non-relativistic matter, 
$a$ is the scale factor of the Universe, and $\sigma_8$ is 
the rms amplitude of over-density at the comoving
8\,$h^{-1}$ Mpc scale ($h$ is the normalized Hubble parameter 
$H_0=100\,h$~km\,sec$^{-1}$Mpc$^{-1}$).
In this paper, we constrain some of the
modified gravity models proposed to solve the dark energy problem,
by using the latest RSD data of 2dFGRS \cite{Percival04},
WiggleZ \cite{Blake}, SDSS LRG \cite{Samushia11}, 
BOSS \cite{Reid12}, and 6dFGRS \cite{Beutler12}.

In this paper, we choose the $f(R)$ gravity and Galileon models as the
representative theories of modified gravity.  A general difficulty of
dark energy models based on modified gravity is the emergence of ``the
fifth force'' that can violate the constraints from local gravity
experiments.  There are a number of mechanisms to suppress the
propagation of the fifth force in local regions: (i) chameleon
mechanism \cite{chame}, (ii) Vainshtein mechanism \cite{Vainshtein},
and (iii) the symmetron mechanism \cite{symmetrons,BBDS}. 
The symmetron mechanism is irrelevant to dark energy
because the energy scale of its simplest potential is too small to 
be used for the late-time cosmic 
acceleration\footnote{Unless the field potential is carefully designed, 
this problem even persists for the 
chameleon field \cite{Brachame}.}.
The $f(R)$ gravity and Galileons are representative models
to explain the acceleration,
avoiding the fifth force problem by mechanisms (i) and (ii),
respectively (see Sec.~\ref{modelsec}, where we give a brief review of
these theories). 

In $f(R)$ gravity several authors put observational bounds on the
parameter $B=(F_{,R}\dot{R}/F)(H/\dot{H})$ by assuming the
$\Lambda$CDM background \cite{Song07,Peiris,Lombriser}, where
$F=\partial f/\partial R$, $F_{,R}=\partial F/\partial R$, and $H$ is
the Hubble parameter (see also Refs.~\cite{fRobsercon} for related
works).  At the level of perturbations the parameter $B$ characterizes
the deviation from the $\Lambda$CDM model.  The joint data analysis of
cluster abundance, the cosmic microwave background (CMB), 
and other observations shows that the value of
$B$ today is constrained to be $B_0<1.1 \times 10^{-3}$ at the 95~\%
confidence level (CL) \cite{Lombriser}.  
The matter and the velocity
power spectra were also computed with $N$-body simulations for the
Hu-Sawicki model (a popular and viable
$f(R)$ dark energy model,
see Eq.~(\ref{fRmo}) below) with $n=1/2$ \cite{Jennings}.
However, those past works did not place explicit
constraints on the viable parameter space of the Hu-Sawicki model from
the RSD data.  

Recently the covariant Galileon dark energy model (whose cosmological
dynamics was studied in Refs.~\cite{GS10,DTPRL,DTPRD}) was confronted
with observations \cite{ApplebyLinder} by using the RSD data of
WiggleZ and BOSS as well as the data of Supernovae Ia (SN Ia), 
CMB, and Baryon Acoustic Oscillations (BAO).  It was found that
this model is severely disfavored over the $\Lambda$CDM.  In this
paper we show that the covariant Galileon is indeed excluded 
at more than $8 \sigma$ CL by using the most recent RSD data of SDSS LRG and
6dFGRS in addition to the WiggleZ and BOSS data. 
We then further
consider the extended Galileon scenario in which a tracker solution
with an arbitrary constant dark energy equation of state smaller than
$-1$ is realized during the matter era.  
Unlike the covariant Galileon
we show that there are some viable parameter spaces compatible with
the current observational data of RSD as well as SN Ia, CMB, and BAO.
 
This paper is organized as follows.  In Sec.~\ref{modelsec} we review
the basic properties of dark energy models based on $f(R)$ gravity and
Galileons.  In Sec.~\ref{persec} we study the
evolution of the growth rate of matter perturbations in these models.
In Sec.~\ref{rsdconsec} we put observational constraints on the
parameters of each model, and Sec.~\ref{consec} is devoted to
conclusions.

\section{Modified Gravity Models}
\label{modelsec}

In general the modified gravity models of dark energy are 
required to recover the Newton gravity at short distances 
for the consistency with local gravity experiments 
in the solar system \cite{Will}.  
As we mentioned in Introduction, there are two known 
mechanisms for the recovery of the Newton gravity 
in local regions.

One is the chameleon mechanism \cite{chame}, 
under which the mass of a scalar degree of freedom is different depending 
on the matter densities in the surrounding environment.
If the effective mass is sufficiently large in the regions of 
high density, the coupling between the field and 
non-relativistic matter can be suppressed 
by having a thin-shell inside a spherically symmetric body.

In $f(R)$ gravity there exists a scalar degree of 
freedom (``scalaron'' \cite{Star80}) with a potential 
of gravitational origin for nonlinear functions $f$ in terms 
of the Ricci scalar $R$.
In this case it is possible to design the scalar potential
such that the chameleon mechanism works at short distances
by choosing appropriate forms of $f(R)$ \cite{fRchame,Capo,Brachame}. 
Explicit $f(R)$ models of dark energy that can satisfy both 
local gravity and cosmological constraints have been 
proposed in Refs.~\cite{HuSa,Star07,Appleby,Tsuji07,Linder}.

Another mechanism for the recovery of Newton gravity 
is the Vainshtein mechanism \cite{Vainshtein}, 
under which nonlinear scalar-field self interactions can
suppress the propagation of the fifth force 
at short distances even in the absence of the field potential.
In the DGP model, a brane-bending mode $\phi$ 
gives rise to the self-interaction of the form $X\square \phi$
(where $X=-(\nabla \phi)^2/2$) through the mixture 
with a transverse graviton \cite{DGPVain1}.
This self-interaction leads to the decoupling of the field $\phi$ from 
matter within a radius much larger than 
the solar-system scale \cite{DGPVain2}.
However this model is plagued by the ghost problem \cite{DGPghost}, 
in addition to the incompatibility with cosmological constraints 
at the background level \cite{DGPcon}.

The field self-interaction $X \square \phi$, which is 
crucial for the recovery of GR in local regions, gives rise to
the field equations invariant under the Galilean shift 
$\phi (x) \to \phi (x)+b_{\mu}x^{\mu}+c$ in 
flat spacetime. 
Nicolis {\it et al.} \cite{Nicolis} derived the general field Lagrangian 
by imposing the Galilean symmetry in Minkowski spacetime.
Deffayet {\it et al.} \cite{Deffayet} obtained the covariant version 
of the Galileon 
Lagrangian keeping the field equations up to second order
in curved backgrounds (see Eq.~(\ref{lag}) below).
In this case the Galilean symmetry is recovered in the limit of 
Minkowski spacetime.
For the covariant Galileon the presence of field self interactions 
other than $X \square \phi$ gives rise to the theoretically allowed 
parameter space in which ghosts and instabilities are 
absent \cite{DTPRL,DTPRD}.
There exists the allowed parameter region 
in which the background cosmological constraints 
are satisfied \cite{Nesseris}.

If we extend the Galileon self-interaction $X \square \phi$ to 
the form $X^{p} \square \phi$ \cite{Kimura}, it gives rise 
to a scenario equivalent to the Dvali-Turner model 
at the background level \cite{DTmodel}.
In this case the dark energy equation of state is 
$w_{\rm DE}=-1-1/(2p-1)$ during the matter era.
However there is an anti-correlation between the large-scale
structure (LSS) and the Integrated-Sachs-Wolfe (ISW) effect in CMB, which 
provides a tight bound on $p$, as $p>4.2 \times 10^{3}$ (95 \%\,CL) \cite{KKY},
in which case the model is practically indistinguishable from 
the $\Lambda$CDM.
This situation is alleviated by taking into account other extended
Galileon terms \cite{DTcondition}. In this extended Galileon scenario there exist
viable parameter spaces in which the LSS and ISW are
positively correlated \cite{DTconstraint}, so that $w_{\rm DE}$ does not 
need to be extremely close to $-1$.

In what follows we review the dark energy models 
based on $f(R)$ gravity, covariant Galileon, and 
extended Galileon, respectively. 

\subsection{$f(R)$ gravity}

The first class is $f(R)$ gravity \cite{fRearly,Star80}
given by the action 
\begin{equation}
S=\frac{M_{\rm pl}^2}{2}\int d^{4}x\sqrt{-g}\,f(R)+S_m\,,
\label{action1}
\end{equation}
where $M_{\rm pl}$ is the reduced Planck mass, $g$ is 
a determinant of the spacetime metric $g_{\mu \nu}$, 
$f(R)$ is a function of the Ricci scalar $R$, and $S_m$
is the action of non-relativistic matter.
We assume that non-relativistic matter 
(with a negligible pressure) is described by 
a barotropic perfect fluid minimally coupled to gravity.
We focus on the metric formalism under which 
the action (\ref{action1}) is varied with respect to $g_{\mu \nu}$
(see Refs.~\cite{Sotiriou,DT10} for reviews).

On the flat Friedmann-Lema\^{i}tre-Robertson-Walker
(FLRW) background with the line-element 
$ds^{2}=-dt^{2}+a^{2}(t)d{\bm{x}}^{2}$ the field 
equations following from the action (\ref{action1}) are \cite{DT10}
\begin{eqnarray}
3F H^2 &=& (FR-f)/2-3H \dot{F}+\rho_m/M_{\rm pl}^2\,,
\label{fRbe1}\\
-2F \dot{H} &=& \ddot{F}-H \dot{F}+\rho_m/M_{\rm pl}^2\,,
\label{fRbe2}
\end{eqnarray}
where $F=\partial f/\partial R$, $H=\dot{a}/a$ is the 
Hubble parameter, $\rho_m$ is the energy density 
of non-relativistic matter, $R=6(2H^2+\dot{H})$, 
and a dot represents a derivative with respect to cosmic time $t$.
{}From these equations we can also
obtain the continuity equation $\dot{\rho}_m+3H\rho_m=0$.
Defining the density parameters 
$\Omega_{\rm DE}=(FR-f)/(6F H^2)-\dot{F}/(HF)$ and 
$\Omega_m=\rho_m/(3F H^2 M_{\rm pl}^2)$, it follows 
that $\Omega_{\rm DE}+\Omega_m=1$ from Eq.~(\ref{fRbe1}).

When we apply $f(R)$ gravity to dark energy the model 
is usually constructed to possess de Sitter solutions 
characterized by $R=R_{\rm dS}=12 H_{\rm dS}^2=$\,constant, 
where $H_{\rm dS}$ is the Hubble parameter at the de Sitter 
solution. Using Eq.~(\ref{fRbe1}) with $\rho_m=0$, it follows that 
\begin{equation}
R_{\rm dS} F(R_{\rm dS})=2f(R_{\rm dS})\,,
\label{decon}
\end{equation}
where $R_{\rm dS}$ is the similar order to
the Hubble parameter $H_0$ today.
Considering the homogeneous perturbations around $R=R_{\rm dS}$, 
the de Sitter solution is stable under 
the following condition \cite{Muller,Faraoni,Amendola}:
\begin{equation}
0<RF_{,R}(R)/F(R) \le 1\,,
\label{dSsta}
\end{equation}
in which case the solutions with 
different initial conditions converge to the
de Sitter attractor.

For the construction of viable $f(R)$ dark energy models
we require the condition $F(R)>0$ to avoid the appearance of 
ghosts \cite{DT10}. 
The mass squared of a scalar degree of freedom
is given by $M^2 \simeq (3F_{,R}(R))^{-1}$ in the regime $M^2 \gg R$, 
so that the condition $F_{,R}(R)>0$ needs to be satisfied to be free from
tachyonic instabilities \cite{Star07}.
In the regions of high density ($\rho \gg H_0^2 M_{\rm pl}^2$)
the mass $M$ needs to be much larger than $H_0$, so that 
the chameleon mechanism can be at work to suppress the 
propagation of the fifth force.

A typical example satisfying the above demands is given by \cite{HuSa}
\begin{equation}
f(R)=R-\lambda R_c \frac{(R/R_c)^{2n}}{(R/R_c)^{2n}+1}\,,
\label{fRmo}
\end{equation}
where $n$, $\lambda$, $R_c$ are positive constants.
There exist other models such as 
$f(R)=R-\lambda R_c [1-(1+R^2/R_c^2)^{-n}]$ ($n>0$) \cite{Star07}
and $f(R)=R-\lambda R_c \tanh (R/R_c)$ \cite{Tsuji07}, 
but the basic properties are similar 
to those in the model (\ref{fRmo})
(the latter corresponds to the limit $n \gg 1$).
The model $f(R)=R-\lambda R_c (R/R_c)^p$ 
($0<p<1$) \cite{Barrow,Amendola} is also cosmologically viable, 
but local gravity constraints put the severe bound 
$p<10^{-10}$ \cite{Capo} 
(i.e., indistinguishable from the $\Lambda$CDM model).
For the model (\ref{fRmo}) the modification of gravity manifests
itself even for $n$ of the order of unity.

In the following we focus on the model (\ref{fRmo}).
The condition (\ref{decon}) for 
the existence of de Sitter solutions is given by 
\begin{equation}
\lambda=\frac{(1+{\cal R}_{\rm dS}^{2n})^2}
{{\cal R}_{\rm dS}^{2n-1} 
(2+2{\cal R}_{\rm dS}^{2n}-2n)}\,,
\label{lam}
\end{equation}
where ${\cal R}_{\rm dS}=R_{\rm dS}/R_c$.
The stability condition (\ref{dSsta}) reads
\begin{equation}
2{\cal R}_{\rm dS}^{4n}-(2n-1)(2n+4)
{\cal R}_{\rm dS}^{2n}+(2n-1)(2n-2) \ge 0\,.
\label{dSsta2}
\end{equation}
For given $n$ this gives the lower bounds on 
${\cal R}_{\rm dS}$ and $\lambda$.
If $n=1$, for example, one has ${\cal R}_{\rm dS} \ge \sqrt{3}$
and $\lambda \ge 8\sqrt{3}/9$.
For $\lambda$ and $n$ of the order of unity, 
Eq.~(\ref{lam}) shows that $R_{\rm dS}$ is roughly of the order 
of $R_c$.
In the limit that $R \gg R_c$ the model 
(\ref{fRmo}) behaves as 
$f(R) \simeq R-\lambda R_c [1+(R/R_c)^{-2n}]$, 
which is close to the $\Lambda$CDM model.

In the regions of high density the chameleon mechanism
works for the model (\ref{fRmo}), in which case 
the fifth force was evaluated in Refs.~\cite{HuSa,Capo}.
{}From the local gravity constraints using the 
violation of weak equivalence principle, 
we obtain the following 
bound \cite{Capo}:
\begin{equation}
n>0.9\,.
\label{ncon}
\end{equation}

Note that $n$ and $\lambda$ are the two fundamental 
parameters for the model (\ref{fRmo}).
For those parameters the ratio 
${\cal R}_{\rm dS}=R_{\rm dS}/R_c$ is known from 
Eq.~(\ref{lam}), which leads to the estimation 
$R_c \approx H_0^2/{\cal R}_{\rm dS}$.
Theoretically natural model parameters are $n$ and 
$\lambda$ of the order of unity, but it remains to 
see whether such parameter space can be consistent 
with the RSD data.

\subsection{Covariant Galileon}

The second class of dark energy models is the covariant 
Galileon \cite{Nicolis,Deffayet}.
This theory is the subclass of the Horndeski's 
general action \cite{Horndeski,DGSZ,KYY11} characterized by 
\be
S=\sum_{i=2}^5\int d^4 x \sqrt{-g}\, {\cal L}_i+S_m\,,
\label{action2}
\ee
where 
\ba
& & {\cal L}_2=K(\phi,X)\,,\quad
{\cal L}_3=-G_3(\phi, X) \square \phi\,, \nonumber \\
& & {\cal L}_4=G_{4}(\phi,X)\, R+G_{4,X}\,[(\Box\phi)^{2}
-(\nabla_{\mu}\nabla_{\nu}\phi)\,(\nabla^{\mu}\nabla^{\nu}\phi)],
\nonumber \\
& & {\cal L}_5=G_{5}(\phi,X)\, G_{\mu\nu}\,(\nabla^{\mu}\nabla^{\nu}\phi) 
\nonumber \\
& &~~~~~~-(G_{5,X}/6)\,[(\Box\phi)^{3}-3(\Box\phi)\,(\nabla_{\mu}
\nabla_{\nu}\phi)\,(\nabla^{\mu}\nabla^{\nu}\phi) \nonumber \\
& &~~~~~~+2(\nabla^{\mu}\nabla_{\alpha}\phi)\,(\nabla^{\alpha}
\nabla_{\beta}\phi)\,(\nabla^{\beta}\nabla_{\mu}\phi)]\,.
\label{lag}
\ea
Here $K$ and $G_i$ ($i=3,4,5$) are functions in terms of 
a scalar field $\phi$ and its kinetic energy 
$X=-g^{\mu \nu} \partial_{\mu} \phi \partial_{\nu} \phi/2$, 
and $G_{\mu\nu}$ is the Einstein tensor.
The covariant Galileon \cite{Deffayet} corresponds to the choice 
\ba
& & K=c_1 M^3 \phi-c_2 X\,,\label{GachoiceK}\\
& & G_3=c_3 X/M^3\,,\\
& & G_4=M_{\rm pl}^2/2-c_4 X^2/M^6\,,\\
& &G_5=3 c_5 X^2/M^9\,,
\label{Gachoice}
\ea
where $c_i$'s are dimensionless constants, and
$M$ is a constant having a dimension of 
mass\footnote{In the original paper of Galileons \cite{Nicolis} 
the linear potential 
$c_1 M^3 \phi$ is expressed as a separate Lagrangian ${\cal L}_1$.
Here we include both the potential and the kinetic term
inside ${\cal L}_2$.}.
Since we are interested in the case where the cosmic acceleration 
is driven by field kinetic terms without a potential, we set 
$c_1=0$ in the following discussion.
Note that the Einstein-Hilbert term $M_{\rm pl}^2 R/2$ appears
inside the Lagrangian ${\cal L}_4$.

On the flat FLRW background the equations of motion 
for the covariant Galileon are \cite{DTPRL}
\begin{eqnarray}
& & 3M_{\rm pl}^2 H^2=\rho_{\rm DE}+\rho_m\,,
\label{basicga1} \\
& & 3M_{\rm pl}^2 H^2+2M_{\rm pl}^2 \dot{H}=-P_{\rm DE}\,,
\label{basicga2}
\end{eqnarray}
where 
\begin{eqnarray}
\rho_{\rm DE} &\equiv& -c_2 \dot{\phi}^2/2
+3c_3 H \dot{\phi}^3/M^3-45 c_4 H^2 \dot{\phi}^4/(2M^6)
\nonumber \\
& &+21c_5 H^3 \dot{\phi}^5/M^9,
\label{rhoga} \\
P_{\rm DE} &\equiv& -c_2 \dot{\phi}^2/2
-c_3 \dot{\phi}^2 \ddot{\phi}/M^3 \nonumber \\
& &+3c_4 \dot{\phi}^3 [8H\ddot{\phi} +(3H^2+2\dot{H})
\dot{\phi}]/(2 M^6) \nonumber \\
& &-3c_5 H \dot{\phi}^4 
[5H \ddot{\phi}+2(H^2+\dot{H})\dot{\phi} ]/M^9\,.
\end{eqnarray}
The dark energy equation of state $w_{\rm DE}$ is 
defined as $w_{\rm DE}=P_{\rm DE}/\rho_{\rm DE}$.

The de Sitter solution ($H=H_{\rm dS}={\rm constant}$) is 
realized for $\dot{\phi}=\dot{\phi}_{\rm dS}={\rm constant}$.
Normalizing the mass $M$ to be $M^3=M_{\rm pl}H_{\rm dS}^2$
and defining $x_{\rm dS} \equiv \dot{\phi}_{\rm dS}/(H_{\rm dS}M_{\rm pl})$,
Eqs.~(\ref{basicga1}) and (\ref{basicga2}) lead to 
the following relations at the de Sitter solution \cite{DTPRL}
\begin{eqnarray}
c_2 x_{\rm dS}^2 &=& 3(2+3\alpha-4\beta)\,,
\label{c2re} \\
c_3 x_{\rm dS}^3 &=& 2+9(\alpha-\beta)\,,
\label{c3re}
\label{dscondition}
\end{eqnarray}
where
\begin{equation}
\alpha \equiv c_4 x_{\rm dS}^4\,,\qquad
\beta \equiv c_5 x_{\rm dS}^5\,.
\label{albe}
\end{equation}
The relations (\ref{c2re}) and (\ref{c3re}) are not 
subject to change under the rescaling 
$x_{\rm dS} \to \gamma x_{\rm dS}$, 
$c_2 \to c_2/\gamma^2$, and $c_3 \to c_3/\gamma^3$, 
where $\gamma$ is a real number.
Hence, the rescaled choices of $c_2$ and $c_3$
lead to the same cosmological dynamics.
This means that we can set $x_{\rm dS}=1$ without 
loss of generality, in which case $c_2$ and $c_3$
are directly known from Eqs.~(\ref{c2re}) and (\ref{c3re}).
It is convenient to use the two model parameters 
$\alpha$ and $\beta$ because the coefficients
of physical quantities can be expressed in terms of them.

We also note that the de Sitter solution given above is
always stable against homogenous 
perturbations \cite{DTPRL,DTPRD}, so the solutions with 
different initial conditions finally converge to the  
de Sitter attractor. 
If we do not demand the existence of de Sitter solutions, 
there are more parameter spaces left for the coefficients 
$c_2$ and $c_3$. 
In such cases, however, the existence of a stable accelerated
attractor is not generally guaranteed.

Introducing the two variables 
$r_1=\dot{\phi}_{\rm dS} H_{\rm dS}/(\dot{\phi} H)$ and 
$r_2=(\dot{\phi}/\dot{\phi}_{\rm dS})^4/r_1$, which satisfy
$r_1=r_2=1$ at the de Sitter solution,
the dark energy density parameter $\Omega_{\rm DE}=
\rho_{\rm DE}/(3M_{\rm pl}^2 H^2)$ can be expressed as 
\begin{eqnarray}
\Omega_{\rm DE}
&=& -(2+3\alpha-4\beta)r_1^3r_2/2+(2+9\alpha-9\beta)r_1^2r_2
\nonumber \\
& &-15\alpha r_1 r_2/2+7\beta r_2\,.
\end{eqnarray}

The autonomous equations for the variables $r_1$ and $r_2$ 
were derived in Ref.~\cite{DTPRD}.
There are three distinct fixed points for this dynamical system: 
(A) $(r_1, r_2)=(0, 0)$, 
(B) $(r_1, r_2)=(1, 0)$, and (C) $(r_1, r_2)=(1, 1)$.
Point (C) is the de Sitter solution,
which is always stable against homogeneous perturbations.
Point (B) corresponds to a tracker 
solution characterized by $\dot{\phi} \propto 1/H$.
For the initial conditions close to the fixed point (A)
the solutions converge to the tracker 
once $r_1$ approaches 1.
The fixed point (B) is followed by the de Sitter point (C)
after $r_2$ grows to the order of 1.
The epoch at which the solutions approach the tracker
depends on the initial values of $r_1$.
For smaller $r_1$ the tracking occurs later.

Along the tracker one has $\Omega_{\rm DE}=r_2$ and 
$w_{\rm DE}=-2/(1+\Omega_{\rm DE})$
after the radiation-dominated epoch.
In the regime close to the fixed point (A) the dark energy 
equation of state is given by $w_{\rm DE}=-1/8$
during the matter era. 
After the solutions reach the tracker, $w_{\rm DE}$ 
changes from $-2$ (matter era) to  $-1$ (de Sitter era).

The joint data analysis of SN Ia, CMB, and BAO shows
that the tracker is disfavored from the 
observational data of the distance measurement \cite{Nesseris}.
For the compatibility with the data the solutions 
need to approach the tracker at late times with 
the minimum value of $w_{\rm DE}$ larger than $-1.3$.
Taking into account the conditions for the avoidance of ghosts and 
Laplacian instabilities \cite{DTPRL,DTPRD}, 
the parameters $\alpha$ and $\beta$ are 
constrained to be 
\begin{equation}
\alpha=1.404 \pm 0.057\,,\qquad
\beta=0.419 \pm 0.023\,,
\label{albe2}
\end{equation}
{}from the joint data analysis of SN Ia (Union2), CMB, and 
BAO \cite{Nesseris}.

We stress that, for given initial conditions of $r_1$ and $r_2$ 
(in other words, given initial conditions of $\dot{\phi}$ and $H$) under
which the solutions enter the tracker at late times, 
there are two parameters 
$\alpha$ and $\beta$ left for the likelihood analysis in Sec.~\ref{rsdconsec}.
The bounds (\ref{albe2}) will be used for such an analysis.

\subsection{Extended Galileon}

The extended Galileon \cite{DTcondition} is described 
by the Horndeski's Lagrangian (\ref{lag}) with the choice
\ba
& & K=-c_2 M_2 ^{4(1-p_2)}X^{p_2}\,,\label{coex1} \\
& & G_3=c_3 M_3^{1-4p_3}X^{p_3}\,,\\
& & G_4=M_{\rm pl}^2/2-c_4 M_4^{2(1-2p_4)}X^{p_4}\,,\\
& & G_5=3 c_5 M_5^{-(1+4p_5)} X^{p_5}\,,\label{coex4}
\label{exGachoice}
\ea
where $c_i$ and $p_i$ ($i=2, \cdots, 5$) are dimensionless constants,
and $M_i$'s are constants having a dimension of mass.  The covariant
Galileon corresponds to $p_2=p_3=1$ and $p_4=p_5=2$.  In
Ref.~\cite{DTcondition} it was shown that there exists a tracker
solution with $H \dot{\phi}^{2q}={\rm constant}$ ($q>0$) for the powers
\ba
& & p_{2}=p\,,\qquad \qquad\,p_{3}=p+(2q-1)/2\,, \nonumber \\
& & p_{4}=p+2q\,,\qquad p_{5}=p+(6q-1)/2\,.
\label{power}
\ea
Note that the tracker corresponds to an attractor 
where the solutions with different 
initial conditions converge to a common trajectory
($p=1$ and $q=1/2$ for the covariant Galileon).

The extended Galileon possesses de Sitter solutions with 
$H=H_{\rm dS}={\rm constant}$ and $\dot{\phi}=\dot{\phi}_{\rm dS}={\rm constant}$.
We relate the masses $M_{i}$'s in Eqs.~(\ref{coex1})--(\ref{coex4}) with $H_{{\rm dS}}$,
as $M_{2}=(H_{{\rm dS}}M_{{\rm pl}})^{1/2}$, 
$M_{3}=(H_{{\rm dS}}^{-2p_{3}}M_{{\rm pl}}^{1-2p_{3}})^{1/(1-4p_{3})}$,
$M_{4}=(H_{{\rm dS}}^{-2p_{4}}M_{{\rm pl}}^{2-2p_{4}})^{1/(2-4p_{4})}$,
and $M_{5}=(H_{{\rm dS}}^{2+2p_{5}}M_{{\rm pl}}^{2p_{5}-1})^{1/(1+4p_{5})}$.
For the existence of de Sitter solutions the following 
relations hold between the coefficients
\ba
\hspace{-0.8cm}& & c_{2}=\frac{3}{2}\left(\frac{2}{x_{{\rm dS}}^{2}}\right)^{p}(2+3\alpha-4\beta),\\
\hspace{-0.8cm}& & c_{3}=\frac{\sqrt{2}}{2p+2q-1}\left(\frac{2}{x_{{\rm dS}}^{2}}\right)^{p+q}
\left[p+3(p+q)(\alpha-\beta)\right],
\label{c3}
\label{powerex}
\ea
where $x_{\rm dS}=\dot{\phi}_{\rm dS}/(H_{\rm dS}M_{\rm pl})$, 
and
\ba
& & \alpha=\frac43 (2p+4q-1) \left( \frac{x_{\rm dS}^2}{2} 
\right)^{p+2q}c_4\,,\\
& & \beta=2\sqrt{2} \left( p+3q-\frac12 \right)
\left( \frac{x_{\rm dS}^2}{2} 
\right)^{p+3q}c_5\,.
\ea

The background equations on the flat FLRW background 
are given by  Eqs.~(\ref{basicga1}) and (\ref{basicga2}), 
with different forms of $\rho_{\rm DE}$ and 
$P_{\rm DE}$ \cite{DTcondition,DTconstraint}.
Introducing the dimensionless variables 
$r_1=(\dot{\phi}_{\rm dS}/\dot{\phi})^{2q} H_{\rm dS}/H$ and 
$r_2=[(\dot{\phi}/\dot{\phi}_{\rm dS})^4/r_1]^{(p+2q)/(1+2q)}$, 
the tracker solution corresponds to $r_1=1$ with the dark 
energy density parameter 
$\Omega_{\rm DE} \equiv \rho_{\rm DE}/(3H^2 M_{\rm pl}^2)=r_2$.
After the radiation era the dark energy equation of state 
$w_{\rm DE}=P_{\rm DE}/\rho_{\rm DE}$ along the tracker
is given by \cite{DTcondition,DTconstraint}
\be
w_{\rm DE}=-\frac{1+s}{1+s\Omega_{\rm DE}}\,,\quad
{\rm where} \quad s=\frac{p}{2q}\,.
\ee
During the matter-dominated epoch ($\Omega_{\rm DE} \ll 1$), 
it follows that $w_{\rm DE} \simeq -1-s$.
The covariant Galileon corresponds to $s=1$ and 
$w_{\rm DE} \simeq -2$, 
in which case the tracker is incompatible with the observational 
data at the background level \cite{Nesseris}.
In the extended Galileon model it is possible to realize $w_{\rm DE}$
close to $-1$ for $s \ll 1$.
The joint data analysis of SN Ia, CMB, and BAO shows that 
the parameter $s$ is constrained to be \cite{DTconstraint}
\be
s=0.034^{+0.327}_{-0.034} \qquad 
(95\, \%~{\rm CL})\,.
\label{eq:s_limit}
\ee
Hence the tracker solution in the range $-1.36<w_{\rm DE}<-1$
can be consistent with the background cosmology.

We choose the initial conditions $r_1=1$ and $r_2=\Omega_{\rm DE} \ll
1$ in the early matter era.  The dark energy density parameter
$\Omega_{\rm DE}^{(0)}$ at the present epoch determines the initial
value of $r_2$ at a given starting redshift.  For given values of $p$
and $q$ there are two parameters $\alpha$ and $\beta$ left for the
likelihood analysis in Sec.~\ref{rsdconsec}.

\section{Cosmological perturbations}
\label{persec}

\subsection{Linear Perturbation Equation}

In Ref.~\cite{DKT} the full linear perturbation equations were 
derived for the most general scalar-tensor
theories with second-order field equations \cite{Horndeski,DGSZ}.
In this section we study the evolution of the cosmic growth rate
in $f(R)$ gravity, covariant Galileon, and extended Galileon.

We consider the scalar metric perturbations $\Psi$ and $\Phi$
in the longitudinal gauge about the flat FLRW background.
The perturbed line element is then given by \cite{Bardeen}
\begin{equation}
ds^{2}=-(1+2\Psi)\, dt^{2}+a^{2}(t)(1+2\Phi) d {\bm x}^2\,.
\label{permet}
\end{equation}
The energy density of non-relativistic matter is decomposed into 
the background and inhomogeneous parts as 
$\rho_m(t)+\delta \rho_m (t, {\bm x})$. 
We write the four-velocity of non-relativistic matter in the 
form $u^{\mu}=(1-\Psi, \nabla^i v)$, where $v$ is the 
rotational-free velocity potential. 
We also introduce the following quantities:
\begin{equation}
\delta \equiv \delta \rho_m/\rho_m\,,\qquad
\theta \equiv \nabla^2 v\,.
\end{equation}
In Fourier space with the comoving wavenumber 
${\bm k}$, the matter perturbation obeys the 
following equations of motion: 
\begin{eqnarray}
& & \dot{\delta}+\theta/a+3\dot{\Phi}=0\,,
\label{per1} \\
& & \dot{\theta}+H \theta-(k^2/a)\Psi=0\,,
\label{per2}
\end{eqnarray}
where $k=|{\bm k}|$.
We introduce the gauge-invariant density contrast
\begin{equation}
\delta_m \equiv \delta+(3aH/k^2)\, \theta\,.
\end{equation}
{}From Eqs.~(\ref{per1}) and (\ref{per2}) 
it follows that
\begin{equation}
\ddot{\delta}_m+2H \dot{\delta}_m
+(k^2/a^2) \Psi=3 ( \ddot{I}
+2H \dot{I})\,,
\label{delmeq0}
\end{equation}
where $I \equiv (aH/k^2)\theta-\Phi$.

In order to confront the models with galaxy clustering surveys, 
we are interested in the modes deep inside the Hubble radius.
In this case we can employ the quasi-static approximation on 
sub-horizon scales, under which the dominant contributions 
to the perturbation equations are those including the terms
$k^2/a^2$, $\delta$, and the mass $M$ of a scalar 
degree of freedom \cite{Star98,Tsuji07}.
Under this approximation we obtain the modified
Poisson equation 
\begin{equation}
(k^2/a^2) \Psi \simeq -4\pi G_{\rm eff} \rho_m \delta\,,
\label{Poisson}
\end{equation}
where $G_{\rm eff}$ is the effective gravitational coupling 
whose explicit form is different depending on the theories \cite{DKT}.
In the framework of General Relativity, $G_{\rm eff}$ is 
equivalent to the gravitational constant $G=1/(8\pi M_{\rm pl}^2)$.

Under the quasi-static approximation on sub-horizon scales, 
the r.h.s. of Eq.~(\ref{delmeq0}) can be neglected relative 
to the l.h.s. of it.
Since $\delta_m \simeq \delta$,
the matter perturbation obeys the following equation: 
\begin{equation}
\ddot{\delta}_m+2H \dot{\delta}_m
-4\pi G_{\rm eff} \rho_m \delta_m \simeq 0\,.
\label{delmeq}
\end{equation}
{}From this equation it is possible to predict 
$\delta_m$, $f_m$, and their evolution, and then
it can be compared with the observed data of
$f_m(z) \sigma_8(z)$. 

\subsection{Comparison with Observations}

The perturbation $\delta_g$ of galaxies is related 
to $\delta_m$ via the bias factor $b$, i.e.,
$\delta_g=b \delta_m$.
The galaxy power spectrum ${\cal P}_g^{s} ({\bm k})$ in 
the redshift space can be modeled as \cite{Kaiser,Tegmark06}
\begin{equation}
{\cal P}_g^{s} ({\bm k})={\cal P}_{gg} ({\bm k})
-2\mu^2 {\cal P}_{g \theta} ({\bm k})
+\mu^4 {\cal P}_{\theta \theta} ({\bm k})\,,
\label{Pred}
\end{equation}
where $\mu={\bm k} \cdot {\bm r}/(kr)$ is the cosine of the angle of
the ${\bm k}$ vector to the line of sight (vector ${\bm r}$).
${\cal P}_{gg} ({\bm k})$ and ${\cal P}_{\theta \theta} ({\bm k})$ are the
real space power spectra of galaxies and $\tilde{\theta}$,
respectively, and ${\cal P}_{g \theta} ({\bm k})$ is the cross power
spectrum of galaxy-$\tilde\theta$ fluctuations in real space, where
$\tilde\theta$ is related to $\theta$ but normalized by the Hubble
velocity, i.e., $\tilde\theta = \theta/(aH)$.  

For the linearly evolving perturbations Eq.~(\ref{per1}) shows that 
$\theta$ is related to the growth rate of matter perturbations, as
\begin{eqnarray}
\tilde\theta &=& \theta/(aH) \simeq -f_m \delta_m = - f_m \delta_g/b\,, \\
f_m &\equiv& \dot{\delta}_m/(H \delta_m)\,,
\label{continuity}
\end{eqnarray}
where we neglected the $\dot{\Phi}$ term.
In this limit we obtain the famous Kaiser's 
formula \cite{Kaiser}:
\begin{eqnarray}
{\cal P}_g^{s} ({\bm k})= (1 + \mu^2 \beta)^2
{\cal P}_{gg} ({\bm k})
\label{kaiserfo}
\end{eqnarray}
where $\beta \equiv f_m/b$ is an anisotropy parameter.  Therefore, the
anisotropy of the measured power spectrum induced by RSD primarily
constrains $\beta$, and we can measure the structure growth rate $f_m$
if $b$ is independently measured. When the independent $b$
measurement is not available, still we can measure the parameter
combination $f_m \sigma_8$ relatively well, because the amplitude
of the observed galaxy power spectrum is proportional to
$(b \sigma_8)^2$, and hence $b$-dependence can be removed
by taking the combination of the two observables in a galaxy
redshift survey, as
$\beta \ ({\cal P}_g^{s})^{1/2} \propto f_m \sigma_8$.

There are some works in which nonlinear corrections, such as
those coming from the velocity distribution of galaxies 
in collapsed structures, are added to the Kaiser's 
formula (\ref{kaiserfo}) \cite{Blake}.
In our work, we focus on the analysis in the linear regime
($k^{-1}>0.2~h^{-1}$~Mpc) where nonlinear 
corrections are small.
However, it is potentially important to fully 
take into account such effects for constraining dark 
energy models more precisely.

For a given model we can numerically solve Eq.~(\ref{delmeq}) to find
$\delta_m(z)$ and $\dot{\delta}_m(z)$ at a given
scale of $x \ h^{-1}$ Mpc.  In GR the cosmic growth rate is
independent of scales for linear perturbations, but in some modified
gravity models like $f(R)$ gravity the scale dependence is present
even at the linear level.  For the models close to the $\Lambda$CDM in the
early cosmological epoch, the evolution of matter perturbations during
the deep matter era is given by $\delta_m \propto a$, which
corresponds to $\dot{\delta}_m=H\delta_m$ and hence $f_m=1$.  The
modification of gravity manifests itself in the late cosmological
epoch.

Now it is easy to predict $f_m(z) \sigma_x(z)$ once the normalization
of $\sigma_x$ is fixed, where $\sigma_x \propto \delta_m$ is the rms
fluctuation amplitude at this scale. This is the quantity that we want
to compare with the measurements of $f_m(z) \sigma_8(z)$.  The
quantity $f_m \sigma_8$ measured from a galaxy redshift survey is
based on $\beta$ and ${\cal P}_g^{s}$, and hence it should essentially
be regarded as $f_m \sigma_x$  with $x$ being the scale
observed by the survey. The conversion into the conventional parameter
$\sigma_8$ is done by assuming the standard shape of the linear matter
power spectrum. Therefore, provided that a given modified gravity
model predicts a matter power spectrum similar to that of $\Lambda$CDM
at $z=0$, we can compare the theoretically predicted $f_m(z)
\sigma_x(z)$ with the observed $f_m(z) \sigma_8(z)$, if we choose $x$
as the relevant galaxy survey scale and normalize $\sigma_x(z)$
as $\sigma_x(z=0) = \sigma_8(z=0)$.  For this normalization, we use
the WMAP7 year bound \cite{Komatsu}
\begin{equation}
\sigma_8 (z=0)=0.811^{+0.030}_{-0.031}\,,
\label{WMAP7}
\end{equation}
which primarily comes from the amplitude at the recombination.
We note that this value is derived by assuming the $\Lambda$CDM model.
Therefore, we also take into account  the following bound 
constrained from clustering of galaxies and galaxy clusters \cite{Rapetti}:
\begin{equation}
\sigma_8 (z=0)=0.795^{+0.030}_{-0.030}\,.
\label{sigma8_clgal}
\end{equation}
In our likelihood analysis we take a slightly wider range 
$0.75< \sigma_8 (z=0) <0.85$ than those constrained above.

The data of $f_m \sigma_8$ used to constrain the models are summarized
in Table \ref{table:fs8constraints}.  We treat these data as
independent measurements, and calculate statistically allowed model
parameter space by the standard $\chi^2$ analysis. This is justified
for different surveys in different volumes, and also approximately for
different redshift bins in one survey, since RSD is mainly measured by
the Fourier modes whose scale is smaller than the redshift bin size.
There is another constraint $f_m \sigma_8=0.49 \pm 0.18$ at $z=0.77$
derived by Guzzo {\it et al.}  \cite{Guzzo08}, but we exclude this
data because of more recent WiggleZ measurement at the redshift
$z=0.78$ \cite{Blake} (see Ref.~\cite{Samushia12}).  

The ranges of scales relevant to these surveys are also given 
in Table \ref{table:fs8constraints}. 
It should be noted that, even though we compare the theory and 
data in terms of $f_m \sigma_8$, we are essentially comparing 
$f_m \sigma_x$, under the condition that the mapping 
between $\sigma_8$ and $\sigma_x$ in modified gravity 
models is similar to that in the concordance cosmological model.

It should be noted that the actual measurements of $f_m \sigma_8$
should be affected by systematic errors coming from the nonlinear
effects, which are ignored in the simple Kaiser formula.  Such
systematic uncertainties are extensively discussed in the experimental
papers deriving $f_m \sigma_8$ such as WiggleZ \cite{Blake} and BOSS
\cite{Reid12}.  It is generally believed that such systematic errors
are smaller than the statistical errors of currently available
surveys, and hence we simply adopt the $f_m \sigma_8$ measurements
reported in the observational papers.

Based on the WMAP7 \cite{Komatsu} constraint on
the today's density parameter of dark energy
\begin{equation}
\Omega_{\rm DE}^{(0)} = 0.725 \pm 0.016\, ,
\label{Omede}
\end{equation}
we adopt $\Omega_{\rm DE}^{(0)} = 0.725$ as the standard value
in this paper, but we also test the dependence on this
parameter within the range of $0.70 \le \Omega_{\rm DE}^{(0)} \le 0.75$,
which is slightly wider than the 1$\sigma$ error of WMAP7.
Below we discuss the evolution of perturbations in three modified
gravity models separately, and compare them with the observed data for
some representative parameter sets. 

\begin{table}[t]
\begin{center}
\begin{tabular}{lcccc}
\hline \hline
$\hspace{8pt}z$ &  $f_m(z) \sigma_8(z)$ & $1/k \ [h^{-1}~\mathrm{Mpc}]$& Survey \\
\hline
0.067 & 0.423$\pm$0.055 & 16 -- 30 & 6dFGRS (2012) \cite{Beutler12} \\
0.17 & 0.51$\pm$0.06 & 6.7 -- 50 & 2dFGRS (2004) \cite{Percival04} \\
0.22 & 0.42$\pm$0.07 & 3.3 -- 50 & WiggleZ (2011) \cite{Blake} \\
0.25 & 0.3512$\pm$0.0583 & 30 -- 200 & SDSS LRG (2011) \cite{Samushia11} \\
0.37 & 0.4602$\pm$0.0378 & 30 -- 200 & SDSS LRG (2011) \cite{Samushia11} \\
0.41 & 0.45$\pm$0.04 & 3.3 -- 50 & WiggleZ (2011) \cite{Blake} \\
0.57 & 0.415$\pm$0.034 & 25 -- 160 & BOSS CMASS (2012) \cite{Reid12} \\
0.6 & 0.43$\pm$0.04 & 3.3 -- 50 & WiggleZ (2011) \cite{Blake} \\
0.78 & 0.38$\pm$0.04 & 3.3 -- 50 & WiggleZ (2011) \cite{Blake} \\
\hline \hline
\end{tabular}
\end{center}
\caption[fs8]{Data of $f_m\sigma_8$ measured from RSD
with the survey references. }
\label{table:fs8constraints}
\end{table}

%
\subsection{$f(R)$ gravity}
%

\begin{figure*}[t]
\centering \noindent
\includegraphics[width=2.3in,height=2.3in]{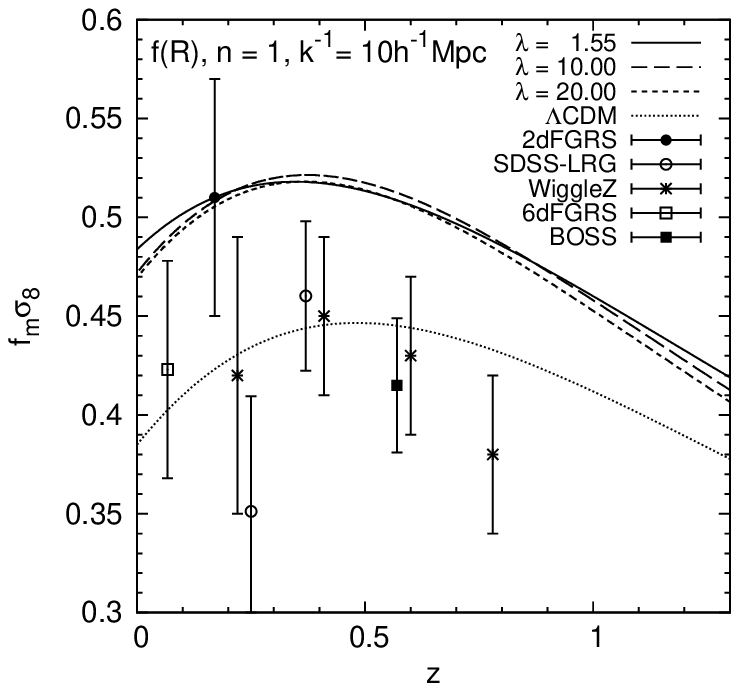} 
\includegraphics[width=2.3in,height=2.3in]{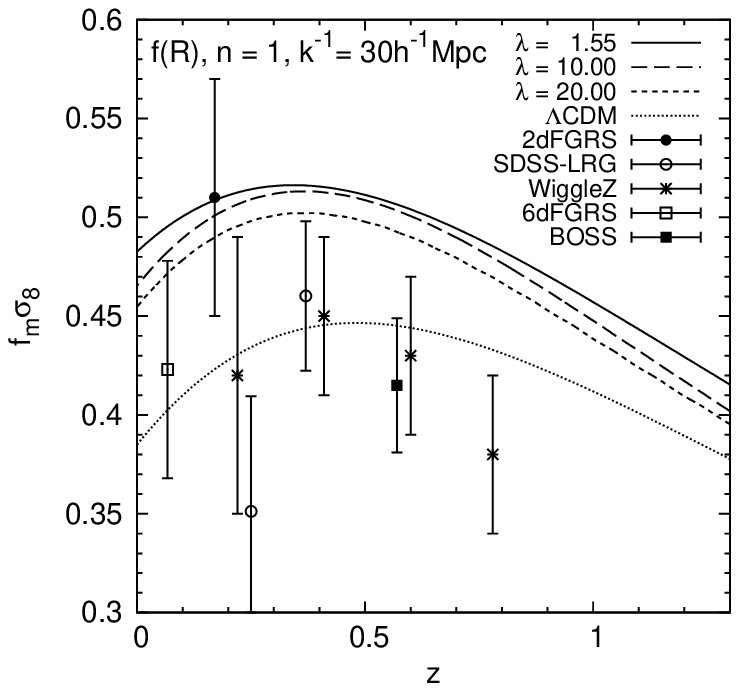}
\includegraphics[width=2.3in,height=2.3in]{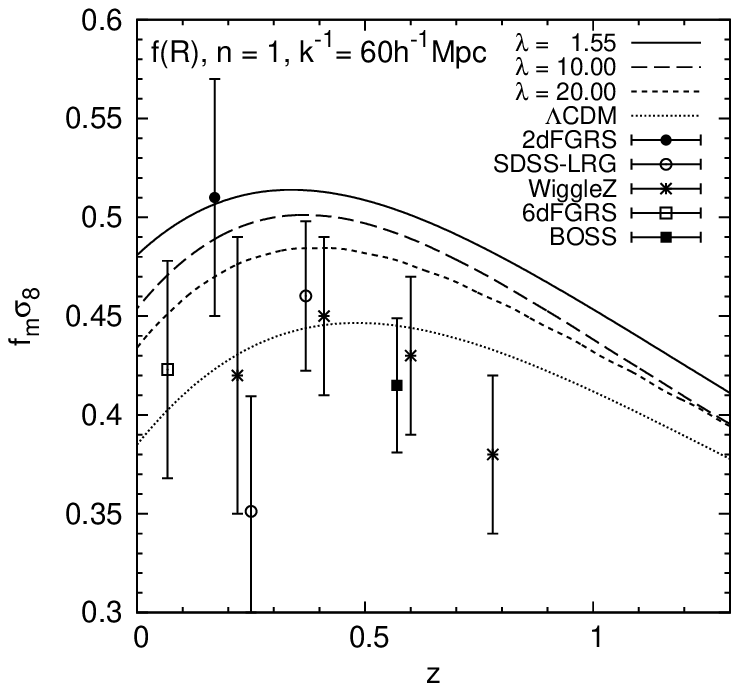} 
\caption{Evolution of $f_m \sigma_8$ versus the redshift $z$
for the $f(R)$ model (\ref{fRmo}) with $n=1$, 
 $\sigma_8 (z=0)=0.811$, and $\Omega_{\rm DE}^{(0)}=0.72$.
The left, middle, and right panels correspond to 
the scales $k^{-1} = 10 \, h^{-1}$, $30 \, h^{-1}$,
$60 \, h^{-1}$~Mpc, respectively.
The solid, long dashed, dashed curves correspond to 
$\lambda=1.55$, $10$, $20$, respectively, whereas the 
dotted curve corresponds to the $\Lambda$CDM.
We also plot the observational bounds on $f_m \sigma_8$
constrained from 2dFGRS \cite{Percival04},
SDSS-LRG \cite{Samushia11}, WiggleZ \cite{Blake},
6dFGRS \cite{Beutler12}, and BOSS \cite{Reid12}.}
 \label{fig1} 
\end{figure*}

\begin{figure*}[t]
\centering \noindent
\includegraphics[width=2.3in,height=2.3in]{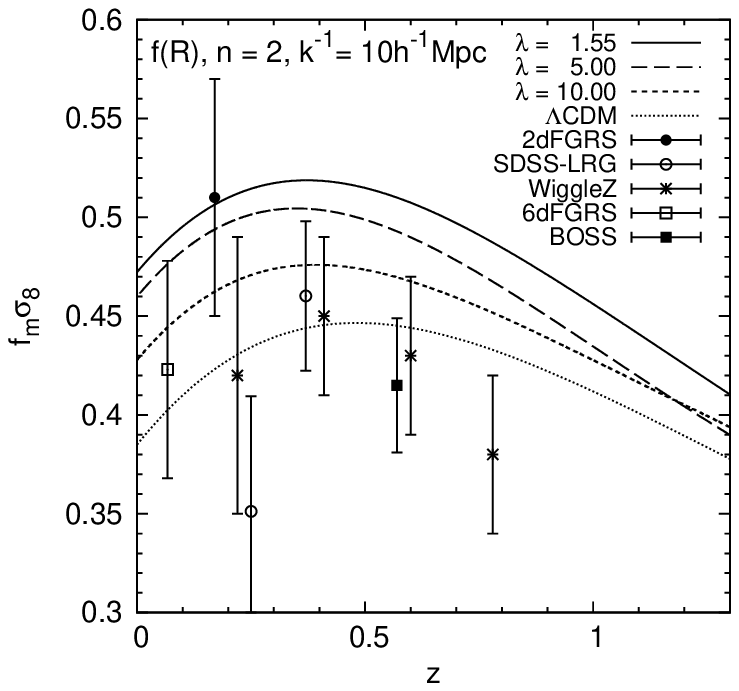} 
\includegraphics[width=2.3in,height=2.3in]{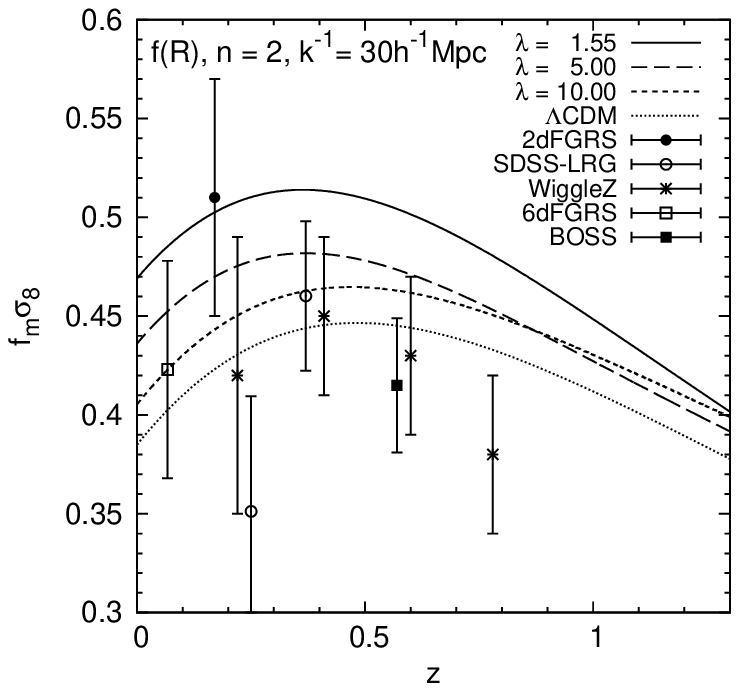}
\includegraphics[width=2.3in,height=2.3in]{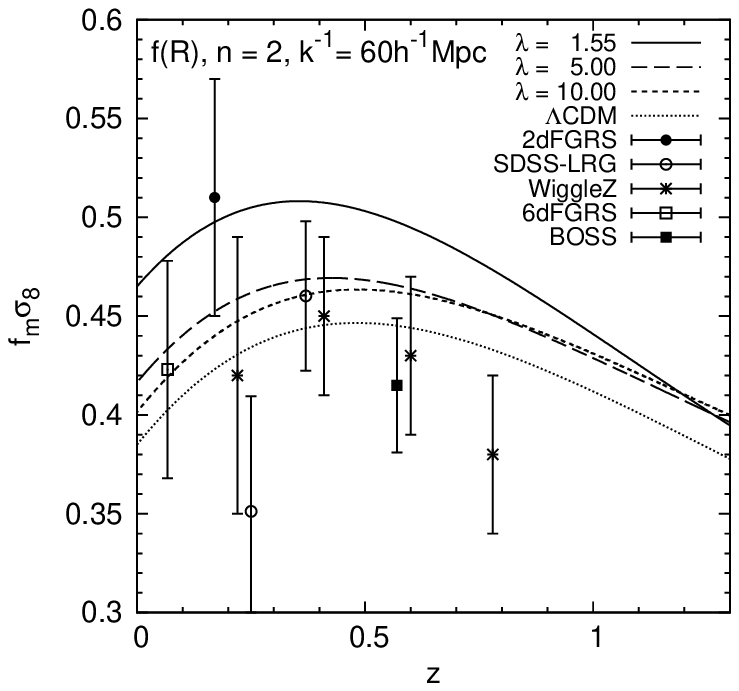} 
\caption{The same as Fig.~\ref{fig1}, but for
$n=2$ and $\lambda = $ 1.55, 5, and 10.
}
 \label{fig2} 
\end{figure*}

The effective gravitational coupling in metric $f(R)$ gravity
is given by \cite{Tsuji07}
\begin{equation}
G_{\rm eff}=\frac{G}{F} 
\frac{1+4m (k/a)^2/R}{1+3m (k/a)^2/R}\,,
\label{GetafR}
\end{equation}
where the parameter $m \equiv RF_{,R}/F$ characterizes
the deviation from the $\Lambda$CDM model ($f=R-2\Lambda$).
In order to avoid ghosts and tachyonic instabilities we require 
$F>0$ and $F_{,R}>0$ (with $R>0$), respectively \cite{Star07}, 
so that $m>0$.

Equation (\ref{GetafR}) shows that the transition from the 
``General Relativistic regime'' ($G_{\rm eff}=G/F$)
to the ``modified gravitational regime'' ($G_{\rm eff}=4G/(3F)$)
occurs around $m \approx (aH/k)^2$.
If such a transition occurs at the redshift $z_c$ larger than 1, 
the transition redshift for the model (\ref{fRmo}) 
can be estimated as \cite{TGMP}
\begin{equation}
z_c \approx \left[ \left( \frac{k}{a_0H_0} \right)^2 
\frac{2n(2n+1)}{\lambda^{2n}} 
\frac{(2\Omega_{\rm DE}^{(0)})^{2n+1}}
{{\Omega_m^{(0)}}^{2(n+1)}} \right]^{1/(6n+4)}-1\,,
\label{zc}
\end{equation}
where $a_0$ is the today's scale factor normalized as
$a_0=1$, $\Omega_{\rm DE}^{(0)}$ and $\Omega_m^{(0)}$
are the today's density parameters of dark energy and non-relativistic
matter respectively. We identify the present epoch 
to be $\Omega_{\rm DE}^{(0)}=0.72$.
The modes with larger $k$ enter the modified gravitational 
regime earlier, so that the growth of perturbations tends to be more 
significant. For smaller $n$ and $\lambda$, the transition 
redshift also gets larger. 
Recall that $n$ and $\lambda$ are bounded from below from 
Eqs.~(\ref{lam})--(\ref{ncon}).

Numerically we solve the perturbation equations without 
using the quasi-static approximation on sub-horizon scales.
The initial conditions of the perturbation $\delta R$ is chosen 
such that the oscillating mode is sub-dominant relative 
to the matter-induced mode 
(along the lines of Refs.~\cite{Star07,Tsuji07}).
The quasi-static approximation based on Eq.~(\ref{delmeq})
with the gravitational coupling (\ref{GetafR}) is accurate 
enough to estimate the growth rate of perturbations 
for sub-horizon modes \cite{Tavakol,Motohashi}.

In Fig.~\ref{fig1} the evolution of $f_m \sigma_8$ is plotted for
$n=1$ and three different values of $\lambda$, with the three different
wavenumbers of $k^{-1} = 10, 30, $ and $60 \ h^{-1}$~Mpc within
the range of galaxy survey scales.
When $\lambda=1.55$
the numerical value of the transition redshift is larger than a few 
and hence the effective gravitational coupling in the regime $z\lesssim 1$ is
given by $G_{\rm eff} \approx 4G/(3F)$.  As we see in Fig.~\ref{fig1},
$f_m \sigma_8$ is larger than that in the $\Lambda$CDM model.  When
$n=1$ and $\lambda=1.55$, the growth rate is almost independent of
scales, which reflects the fact that $z_c$ is larger than 1 on those
scales.

In Fig.~\ref{fig1} we find that the dispersion of $f_m \sigma_8$
with respect to $\lambda$ appears on larger scales.
This comes from the fact that, for smaller $k$ and larger $\lambda$, 
the transition redshift (\ref{zc}) gets smaller.
Even for the scale $60\,h^{-1}$\,Mpc and $\lambda=20$, however, 
$f_m \sigma_8$ is larger than that in the $\Lambda$CDM, 
so that the $f(R)$ model with $n=1$ can be distinguished from 
the $\Lambda$CDM.

For larger $n$ the transition redshift $z_c$ tends to be smaller, 
and the growth rate of $\delta_m$ today gets 
smaller for given $k$ and $\lambda$.
In Fig.~\ref{fig2} we show the evolution of $f_m \sigma_8$
for $n=2$ and $\lambda=$1.55, 5, and 10.
In this case the dispersion of $f_m \sigma_8$ with respect 
to $\lambda$ appears even for the scale 
$10\,h^{-1}$~Mpc.
On the scales larger than $30\,h^{-1}$~Mpc the values of 
$f_m \sigma_8$ are almost degenerate for $\lambda \gtrsim 10$, 
whereas for $\lambda={\cal O}(1)$ the growth rate 
is still large even on the scale $60\,h^{-1}$~Mpc.

When compared with the observed data in Figs.~\ref{fig1}
and \ref{fig2}, the $f(R)$
model (\ref{fRmo}) with $n$ and $\lambda$ of the order of 1 is in
tension with most of the data.  In Sec.~\ref{rsdconsec} we place
observational constraints on the model parameters 
$n$ and $\lambda$.

\subsection{Covariant Galileon}

The effective gravitational coupling for the covariant Galileon 
is given by  \cite{Kase} (see also Refs.~\cite{DKT,ApplebyJCAP,Barreira12,Barreira})
\begin{equation}
G_{\rm eff}=\frac{2M_{\rm pl}^2 
({\cal C}_4^2-{\cal C}_3 {\cal C}_5)}
{2{\cal C}_1 {\cal C}_2 {\cal C}_4-{\cal C}_1^2 {\cal C}_5
-{\cal C}_2^2 {\cal C}_3}G\,,
\label{GGa}
\end{equation}
where the time-dependent coefficients ${\cal C}_i$'s are 
\begin{eqnarray}
\hspace{-0.6cm}{\cal C}_1 &=& 2M_{\rm pl}^2+3c_4 \dot{\phi}^4/M^6
-6 c_5 H \dot{\phi}^5/M^9,\\
\hspace{-0.6cm}{\cal C}_2 &=& -c_3 \dot{\phi}^2/M^3+12 c_4 H\dot{\phi}^3/M^6
-15c_5 H^2 \dot{\phi}^4/M^9, \\
\hspace{-0.6cm}{\cal C}_3 &=& 2M_{\rm pl}^2-c_4 \dot{\phi}^4/M^6
-6c_5 \dot{\phi}^4 \ddot{\phi}/M^9,\\
\hspace{-0.6cm}{\cal C}_4&=&
12c_4 \dot{\phi}^2 \ddot{\phi}/M^6+4c_4 H\dot{\phi}^3/M^6
-6c_5 \dot{H}\dot{\phi}^4/M^9\nonumber \\
& &-6c_5 H^2 \dot{\phi}^4/M^9
-24 c_5 H \dot{\phi}^3 \ddot{\phi}/M^9, 
\end{eqnarray}
\begin{eqnarray}
\hspace{-0.5cm}{\cal C}_5&=&c_2-4c_3 H \dot{\phi}/M^3
-2c_3 \ddot{\phi}/M^3+26c_4 H^2 \dot{\phi}^2/M^6 \nonumber \\
& &+12c_4 \dot{H} \dot{\phi}^2/M^6
+24c_4 H \dot{\phi} \ddot{\phi}/M^6  \nonumber \\
& &-36c_5 H^2 \dot{\phi}^2 \ddot{\phi}/M^9 -24c_5H \dot{\phi}^3 (H^2 
+\dot{H})/M^9.
\end{eqnarray}
%

\begin{figure}
\centering \noindent
\includegraphics[width=3.3in,height=3.3in]{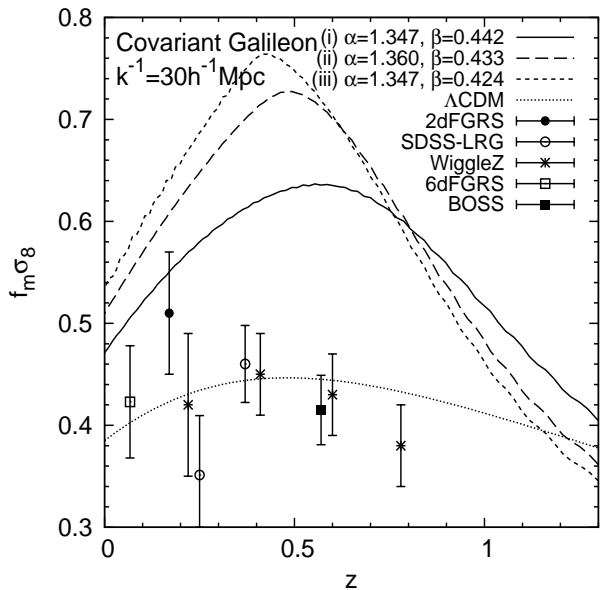} 
\caption{Evolution of $f_m \sigma_8$ versus the redshift $z$
for the scale $30\,h^{-1}$\,Mpc in the covariant Galileon model 
with $\sigma_8 (z=0)=0.811$ and $\Omega_{\rm DE}^{(0)}=0.73$.
The solid, long dashed, and dashed curves 
correspond to the cases 
(i) $\alpha=1.347$, $\beta=0.442$,
(ii) $\alpha=1.360$, $\beta=0.433$, and
(iii) $\alpha=1.347$, $\beta=0.424$, respectively, 
whereas the dotted curve shows the evolution 
of $f_m \sigma_8$ in the $\Lambda$CDM model.
The observational data are the same as those plotted 
in Fig.~\ref{fig1}.}
 \label{fig3} 
\end{figure}

At the de Sitter solution ($r_1=r_2=1$) 
the formula (\ref{GGa}) reduces to 
\begin{equation}
\frac{G_{\rm eff}}{G}=
\frac{1}{3(\alpha-2\beta)}\,.
\label{Geffde}
\end{equation}
This shows that $\alpha-2\beta>0$ to avoid ghosts.
The late-time tracker corresponds to the solution
along which the conditions $r_1 \ll 1$ and $r_2 \ll 1$ are
satisfied during most of the cosmological epoch prior to 
the dominance of dark energy.
In this regime, expansion of $G_{\rm eff}$ around 
$r_1=0$ and $r_2=0$ gives
\begin{equation}
\frac{G_{\rm eff}}{G} \simeq 
1+\left( \frac{255}{8}\beta+\frac{211}{16} \alpha r_1 \right)r_2\,.
\end{equation}
The non-ghost condition corresponds to 
$\beta r_2>0$ \cite{DTPRL}, so that $G_{\rm eff}>G$.
Hence the growth rate of $\delta_m$ is larger than that
in the $\Lambda$CDM.

Numerically we solve the full perturbation equations presented 
in Refs.~\cite{Kase,DKT} without using 
the quasi-static approximation for the model parameters 
given in Eq.~(\ref{albe2}).
The quantity $r_1$ needs to be much smaller than 1 during the early 
matter era, in which case the solutions approach the tracker at late times.
The initial conditions for the perturbations are chosen to 
satisfy $\dot{\Phi}=0$ and $\dot{\delta \phi}=0$ in the 
deep matter era (see Ref.~\cite{Kase} for detail).

In order to avoid ghosts and Laplacian instabilities 
the parameters $\alpha$ and $\beta$ are restricted to be 
in the range $0<\alpha-2\beta<2/3$ \cite{DTPRL,DTPRD}.
Numerically we find that, for $\alpha-2\beta>0.505$, 
the perturbations show violent instabilities during the transition 
from the matter era to the de Sitter epoch.
This is associated with the fact that the effective gravitational 
coupling (\ref{GGa}) tends to cross 0 for larger $\alpha$.
Hence we focus on the parameter space $0<\alpha-2\beta<0.505$,
in addition to the constraints from the background
evolution [Eq.~(\ref{albe2})].

In Fig.~\ref{fig3} we plot the evolution of $f_m \sigma_8$ for the
scale $k^{-1} = 30\,h^{-1}$\,Mpc with several different values of $\alpha$ and
$\beta$.  For larger values of $\alpha-2\beta$ the growth of
$\delta_m$ tends to be more significant, so that $f_m \sigma_8$ gets
larger. Even in case (i) (see caption of Fig.~\ref{fig3}), which
gives the smallest $f_m \sigma_8$ within the allowed ranges of
$\alpha$ and $\beta$, the model is in tension with
the recent RSD data because of the large growth rate of
$\delta_m$.  Numerically we also confirm that the evolution of $f_m
\sigma_8$ is practically independent of the scales in the range
$10\,h^{-1}$~Mpc $\lesssim\,k^{-1} \lesssim$ $60\,h^{-1}$~Mpc.

Since the growth rate of linear perturbations is practically 
independent of scales, only the total amplitude of the matter 
power spectrum is subject to change relative to that 
in the $\Lambda$CDM.
The matter power spectra of covariant Galileon and $\Lambda$CDM models
were explicitly evaluated in Fig.~6 of Ref.~\cite{Barreira12}, 
which shows the property mentioned above.
Hence the mapping between $\sigma_8$ and $\sigma_x$
to derive the constraint on $f_m \sigma_8$ in the $\Lambda$CDM
model should be valid for the covariant Galileon model as well.

\subsection{Extended Galileon}
%

\begin{figure}
\begin{centering}
\includegraphics[width=3.5in,height=3.0in,clip,trim=38 0 26 0]{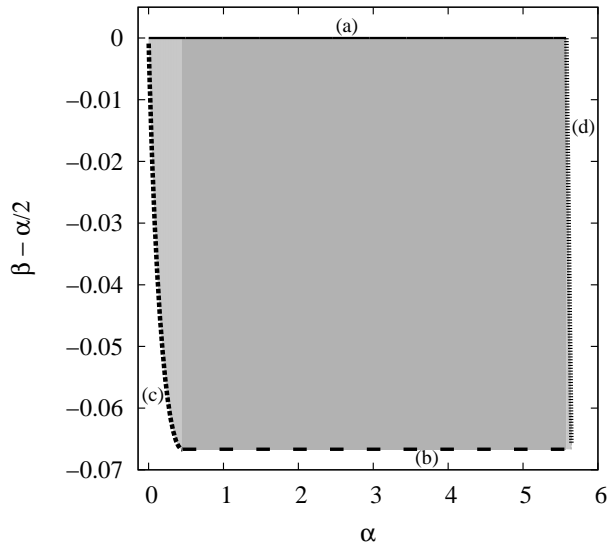} 
\par\end{centering}
\caption{The grey region shows the theoretically allowed parameter
  space for the extended Galileon model with $p=1$ and $q=5/2$ (the
  region in which ghosts and Laplacian instabilities are absent).
  Each border corresponds to (a) $\beta=\alpha/2$, (b)
  $\beta=\alpha/2-1/15$, (c)
  $\beta=(408\alpha+68-2\sqrt{17}\sqrt{3(272-75\alpha)\alpha+68})/561$,
  and (d) $\beta=(242-15\alpha+4\sqrt{3630-495\alpha})/99$,
  respectively. Taken from Ref.~\cite{DTconstraint}.}
\centering{} \label{fig4} 
\end{figure}

\begin{figure}
\centering \noindent
\includegraphics[width=3.3in,height=3.3in]{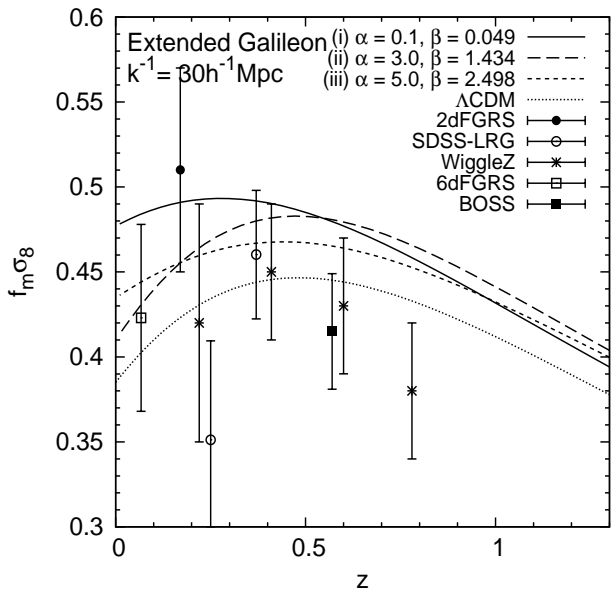} 
\caption{Evolution of $f_m \sigma_8$ versus the redshift $z$
for the scale $30\,h^{-1}$\,Mpc in the extended Galileon model 
with $\sigma_8 (z=0)=0.811$ and $\Omega_{\rm DE}^{(0)}=0.72$.
The solid, long dashed, and dashed curves 
correspond to cases 
(i) $\alpha=0.1$, $\beta=0.049$,
(ii) $\alpha=3$, $\beta=1.434$, and
(iii) $\alpha=5$, $\beta=2.498$, respectively, 
whereas the dotted curve corresponds to 
the $\Lambda$CDM.
The observational data are the same as those plotted 
in Fig.~\ref{fig1}.}
 \label{fig5} 
\end{figure}

The effective gravitational coupling of for the extended Galileon
is given by the form (\ref{GGa}) with more complicated coefficients 
${\cal C}_i$ (see Ref.~\cite{DTconstraint} for details).
At the de Sitter solution ($r_1=r_2=1$) it follows that 
\begin{equation}
\frac{G_{\rm eff}}{G}=
\frac{2}{2(1-p)+3(1+2q)(\alpha-2\beta)}\,.
\label{Geffextended}
\end{equation}
The general expression of $G_{\rm eff}$ is quite involved even along
the tracker ($r_1=1$).  In the following we focus on the tracker
solution ($r_1=1$) for the model $p=1$ and $q=5/2$, in which case
$s=0.2$ and hence $w_{\rm DE}=-1.2$ during the matter era.  
This model shows some deviation from the $\Lambda$CDM model, but
it is still consistent with the background-level observations
(see Eq.~(\ref{eq:s_limit})).
In the regime $r_2 \ll 1$, $G_{\rm eff}$ is
approximately given by
\ba
\frac{G_{\rm eff}}{G} &\simeq&
1+[48411\alpha^{2}-3\alpha(3560+60291\beta) \nonumber \\
& &+22(50+924\beta+7641\beta^{2})]r_2 \nonumber \\
& &/[10(308-1536\alpha+3201\beta)]\,,
\label{Geffextended2}
\ea
whereas at the de Sitter solution $G_{\rm eff}/G=1/[9(\alpha-2\beta)]$.

In Refs.~\cite{DTcondition,DTconstraint} the authors clarified the 
parameter space in which ghosts and Laplacian instabilities 
of scalar and tensor perturbations are absent.
Figure \ref{fig4} shows such a viable region, which is surrounded
by the 4 borders given in the caption.
On the line (a) $\beta=\alpha/2$, one has $G_{\rm eff}/G \to \infty$
at the de Sitter solution and $G_{\rm eff}/G \simeq 
1+4(275-129 \alpha)r_2/[5(616+129 \alpha)]$ for
$r_2 \ll 1$. In the latter regime $G_{\rm eff}$ gets
larger for smaller $\alpha$.
Hence today's values of $f_m \sigma_8$ on the line (a) are 
large around the region $\alpha=\beta=0$.
The case (i) in Fig.~\ref{fig5} corresponds to 
such an example.

On the line (b) shown in Fig.~\ref{fig4}, we have $G_{\rm eff}/G=5/6$
at the de Sitter solution and $G_{\rm eff}/G \simeq 
1+13r_2/25$ for $r_2 \ll 1$. 
In the regime $r_2 \ll 1$, $G_{\rm eff}$ on the line (b) 
is larger than that on the line (a) for the same value of $\alpha$, 
but at the de Sitter solution $G_{\rm eff}$ is even smaller
than that for the $\Lambda$CDM.
The case (ii) in Fig.~\ref{fig5} corresponds to such 
an example, which exhibits rapid decrease of 
$f_m \sigma_8$ for $z<0.5$.
This case shows a better compatibility with the RSD data 
in the low redshift regime relative to the case (i).

Around the line (a), the effective gravitational coupling $G_{\rm eff}$ 
in the regime $r_2 \ll 1$ is smaller than $G$ for $\alpha>275/129$.
Although $G_{\rm eff}$ goes to infinity at the de Sitter solution, 
the growth rates of matter perturbations are
suppressed in high redshifts for $\alpha>275/129$.
The case (iii) in Fig.~\ref{fig5} corresponds to such an example, 
which is consistent with some of the RSD data
because of the suppression of $f_m \sigma_8$ in the 
early cosmological epoch.

We also confirm that the evolution of $f_m \sigma_8$ is practically 
independent of the scales in the range
$10\,h^{-1}$~Mpc $\lesssim\,k^{-1} \lesssim$ $60\,h^{-1}$~Mpc.

\section{Observational constraints from RSD}
\label{rsdconsec}

Here we determine quantitative model parameter space that is excluded
from the RSD data for the three modified gravity models discussed in
Sec.~\ref{persec}. The confidence regions of excluded parameter
spaces are calculated by the standard $\chi^2$ analysis.  For a
given set of model parameters, $\chi^2_{\rm obs}$ is calculated from
the 9 data points of $f_m(z) \sigma_8(z)$, and the exclusion
confidence level at this point is the probability of 
$\chi^2 <\chi^2_{\rm obs}$ for the $\chi^2$ distribution with $d$ degrees of
freedom, where $d = N_d - N_m$, $N_d = 9$ is the data number, and
$N_m$ is the number of model parameters. For all the three modified
gravity models considered here, $N_m = 2$. 

In three classes of modified gravity models studied below, 
we first carry out the likelihood analysis by fixing $\sigma_8 (z=0)=0.811$.
We then discuss the cases in which $\sigma_8 (z=0)$ is varied 
in the range $0.75<\sigma_8 (z=0)<0.85$.
We also fix the value of $\Omega_{\rm DE}^{(0)}$ to be 0.725 constrained 
by the background cosmology first and then check whether the 
constraints are subject to change by varying $\Omega_{\rm DE}^{(0)}$
in the range $0.70<\Omega_{\rm DE}^{(0)}<0.75$.

\subsection{$f(R)$ model}
%

\begin{figure*}[t]
\centering \noindent
\includegraphics[width=2.3in,height=2.3in]{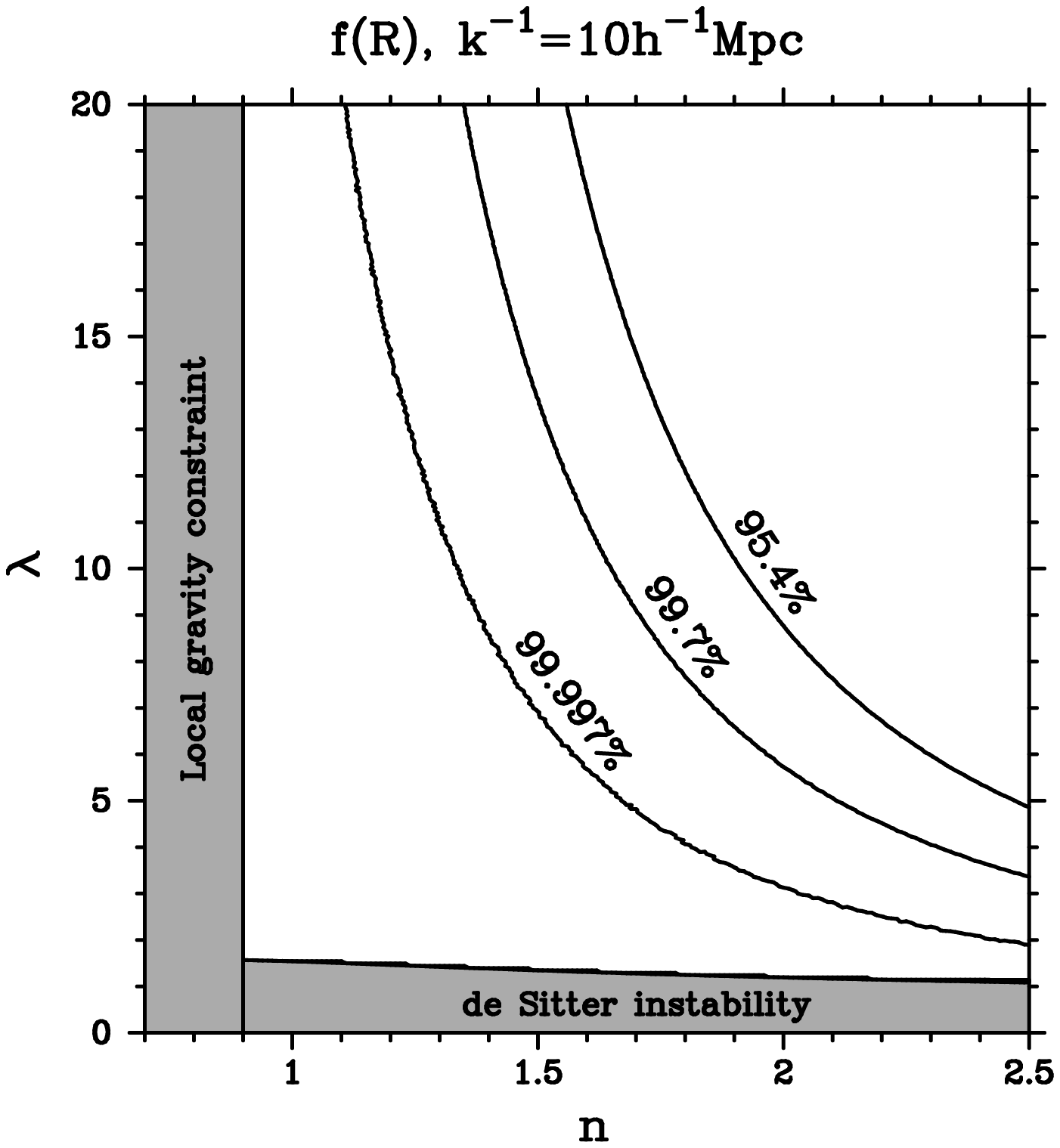} 
\includegraphics[width=2.3in,height=2.3in]{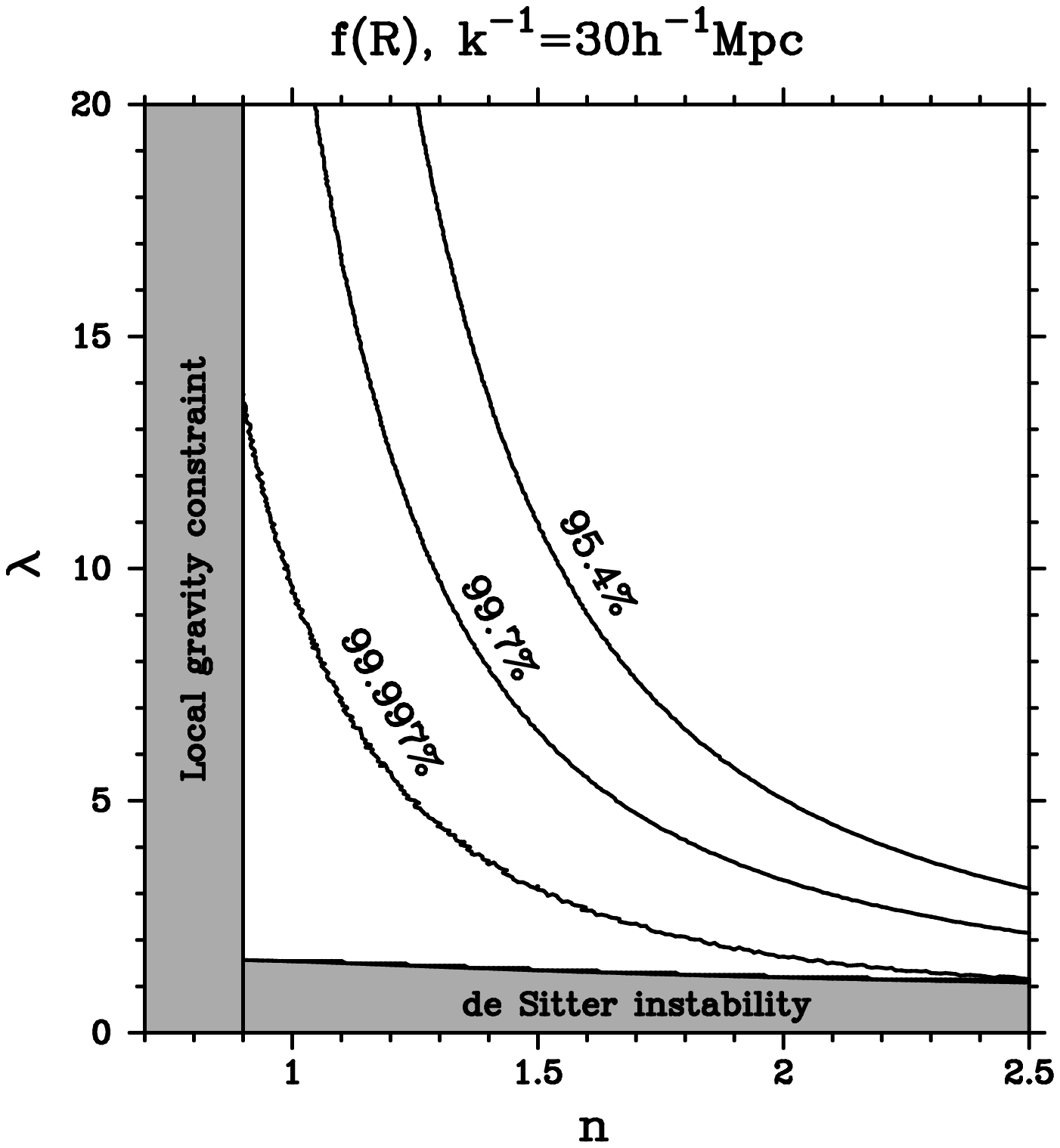}
\includegraphics[width=2.3in,height=2.3in]{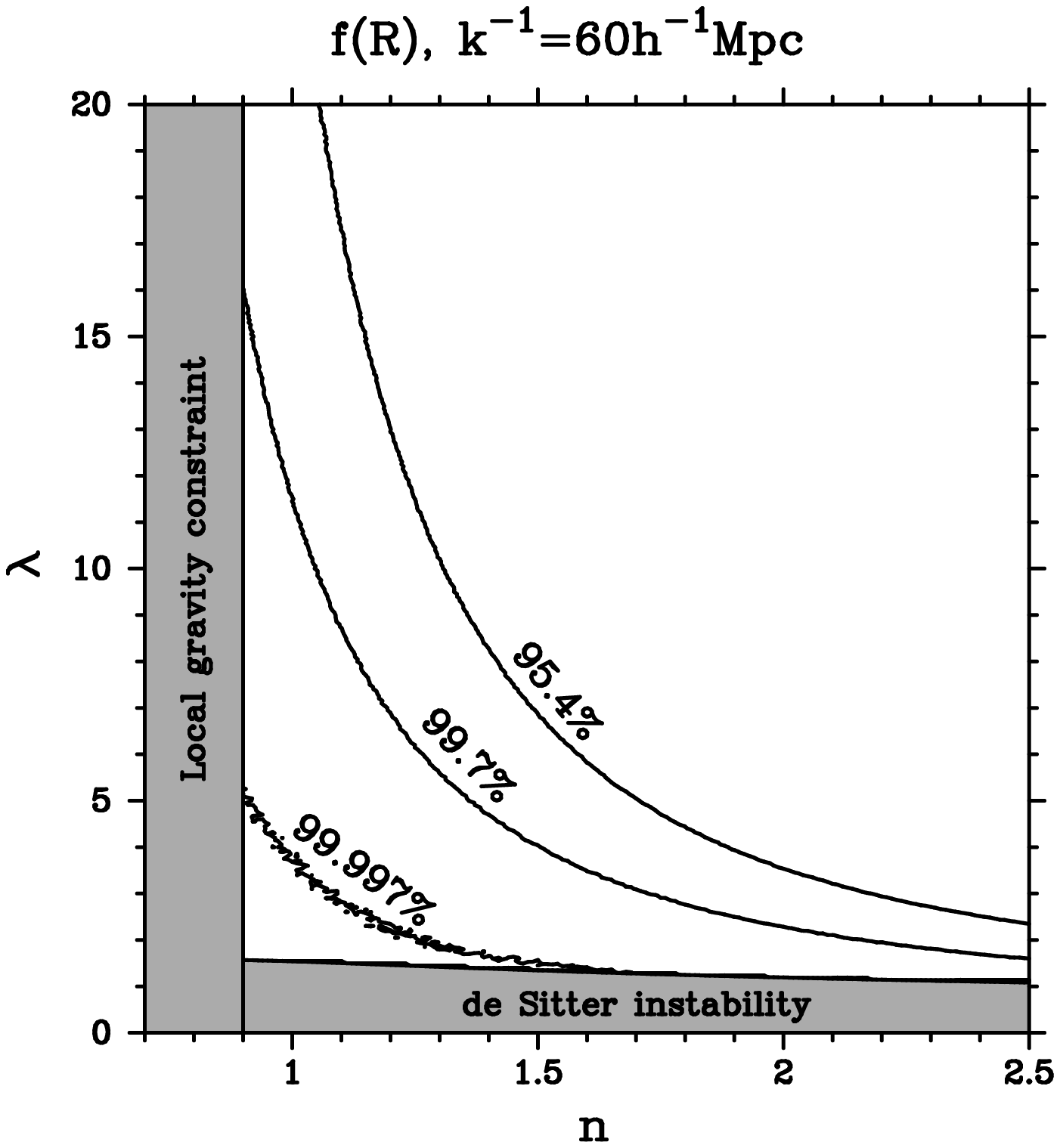} 
\caption{Constraints on the Hu-Sawicki
$f(R)$ gravity model (\ref{fRmo}) from the RSD data 
in the $(n,\lambda)$ plane 
with $\sigma_8 (z=0)=0.811$ and $\Omega_{\rm DE}^{(0)}=0.72$
for three different scales of $k^{-1} = 10 \, h^{-1}$ Mpc (left), 
$30 \, h^{-1}$ (middle), $60 \, h^{-1}$~Mpc (right). 
The lower-left regions are excluded by the confidence
levels indicated along with the contours.
}
 \label{fig6} 
\end{figure*}

We first place observational constraints on the $f(R)$ model
(\ref{fRmo}) for three different scales $10 \, h^{-1}$, $30 \,
h^{-1}$, and $60 \, h^{-1}$~Mpc.  We identify the present epoch to be
$\Omega_{\rm DE}^{(0)}=0.72$ with the amplitude $\sigma_8 (z=0)=0.811$. 
In Fig.~\ref{fig6} we plot the
parameter space in the $(n, \lambda)$ plane constrained by the RSD
data shown in Table \ref{table:fs8constraints}.  The grey regions are
forbidden by the violations of the de Sitter stability condition
(\ref{dSsta2}) as well as that of the local gravity constraint
(\ref{ncon}).  Figure \ref{fig6} shows that observational constraints
tend to be tighter for smaller scales.

The model with $n=1$ and $\lambda<20$ is outside the 95.4 \% CL
contour for the scales $10 \, h^{-1} {\rm Mpc} \lesssim k^{-1}
\lesssim 60 \, h^{-1}$~Mpc.  This reflects the fact that the cosmic
growth rates in this parameter space are too large to be compatible
with the recent RSD data in the regime $z \lesssim 1$, irrespective of
the values of $\lambda$ (see Fig.~\ref{fig1}).

For $n=2$, the parameter $\lambda$ is constrained to be $\lambda>8$
($10 \, h^{-1} {\rm Mpc}$), $\lambda>5$ ($30 \, h^{-1} {\rm Mpc}$),
and $\lambda>3$ ($60 \, h^{-1} {\rm Mpc}$) at 95.4 \% CL.  In terms of
the dimensionless parameter $m=RF_{,R}(R)/F(R)$ characterizing the
deviation from the $\Lambda$CDM \cite{Amendola}, we have $m(z=0)=9.6
\times 10^{-5}$ for $n=2, \lambda=8$ and $m(z=0)=6.3 \times 10^{-4}$
for $n=2, \lambda=5$.  Hence, for $n=2$, today's values of $m$ are
bounded to be $m(z=0)<9.6 \times 10^{-5}$ ($10 \, h^{-1} {\rm Mpc}$)
and $m(z=0)<6.3 \times 10^{-4}$ ($30 \, h^{-1} {\rm Mpc}$) at 95.4 \%
CL.  The constraint on $m(z=0)$ coming from the integrated Sachs-Wolfe
effect of the CMB anisotropies is much weaker, $m(z=0) \lesssim 1$
\cite{Peiris,fRde}.  This shows that the measurement of the cosmic
growth rate from RSD is sufficiently powerful to constrain the $f(R)$
gravity model tightly.

If we choose the values of $n$ greater than 3, the 
model is close to the $\Lambda$CDM. 
In this case, however, the scalaron mass is very heavy in the early 
cosmological epoch, so that there is a severe fine tuning 
for the initial conditions of perturbations to avoid the dominance
of the oscillating mode \cite{Star07,Tsuji07}. 
Hence it is not so meaningful to consider
such cases even if $\chi^2$ may be reduced relative to the case
$n \lesssim 2.5$.  

In addition to two-dimensional constraints, we carry out the 
likelihood analysis by fixing one of the
parameters $n$ and $\lambda$.
The results of such one-dimensional constraints are 
summarized in Table \ref{table:1Dconstraints}. 
For larger values of $n$ (or $\lambda$), the allowed region
of the other parameter tends to be wider.

\begin{table}[t]
\begin{center}
\begin{tabular}{ccc}
\hline \hline
Theory & Fixed parameter & Constraint \\
\hline
$f(R)$& $n=1.0$ & $\lambda \gg 20$\\
& $n=1.5$ & $\lambda > 9.7$\\
& $n=2.0$ & $\lambda > 4.6$\\
& $\lambda = 1.55$ & $n \gg 2.5$\\
& $\lambda = 5.0$ & $n > 1.92$\\
& $\lambda = 10$ & $n > 1.48$\\
Extended & $\alpha = 0.1$ & $-0.061 < \beta-\alpha /2 < -0.039$\\
Galileon & $\alpha = 3.0$ & $-0.063 < \beta-\alpha /2 < 0$\\
($s=0.2$)& $\alpha = 5.0$ & $-0.063 < \beta-\alpha /2 < 0$\\
& $\beta-\alpha /2 = -0.01$ & $\alpha > 1.0$\\
& $\beta-\alpha /2 = -0.02$ & $\alpha > 0.7$\\
& $\beta-\alpha /2 = -0.066$ & no allowed region\\
\hline \hline
\end{tabular}
\end{center}
\caption{One-dimensional constraints on $f(R)$ and extended Galileon 
theories (95.4~\% CL). The other parameters are fixed to be 
$k^{-1} = 30\,h^{-1}$~Mpc, $\sigma_8(z=0) 
= 0.811$, and $\Omega_{\rm DE}^{(0)} = 0.72$.
``$\gg$" means that there are no allowed regions 
within our calculation range.  
Covariant Galileon theory has no allowed parameter space 
at 95.4~\% CL.}
\label{table:1Dconstraints}
\end{table}

We also study how the two-dimensional constraints are subject to change 
when $\sigma_8(z=0)$ and $\Omega_{\rm DE}^{(0)}$ vary. 
For larger $\sigma_8(z=0)$ and smaller $\Omega_{\rm DE}^{(0)}$  
the observational contours shift the region of smaller values of 
$n$ and $\lambda$, so that the constraints get tighter. 
We confirm that the essential result does not change
by varying these parameters in the range 
$0.75<\sigma_8(z=0)<0.85$ and $0.70<\Omega_{\rm DE}^{(0)}<0.75$.

In Refs.~\cite{Jennings} it was pointed out that
non-linear perturbations in $f(R)$ gravity start to provide 
some contribution to the redshift-space power spectrum even 
for $k=0.02$\,$h$\,Mpc$^{-1}$ (see Fig.~4 of \cite{Jennings}).
The non-linear effect suppresses
the redshift-space power spectrum especially for 
$k \gtrsim 0.1$\,$h$\,Mpc$^{-1}$.
It remains to be seen how the observational constraints
are subject to change by non-linear perturbations.

\subsection{Covariant Galileon}
%

\begin{figure}
\begin{centering}
\includegraphics[width=3.4in,height=3.4in]{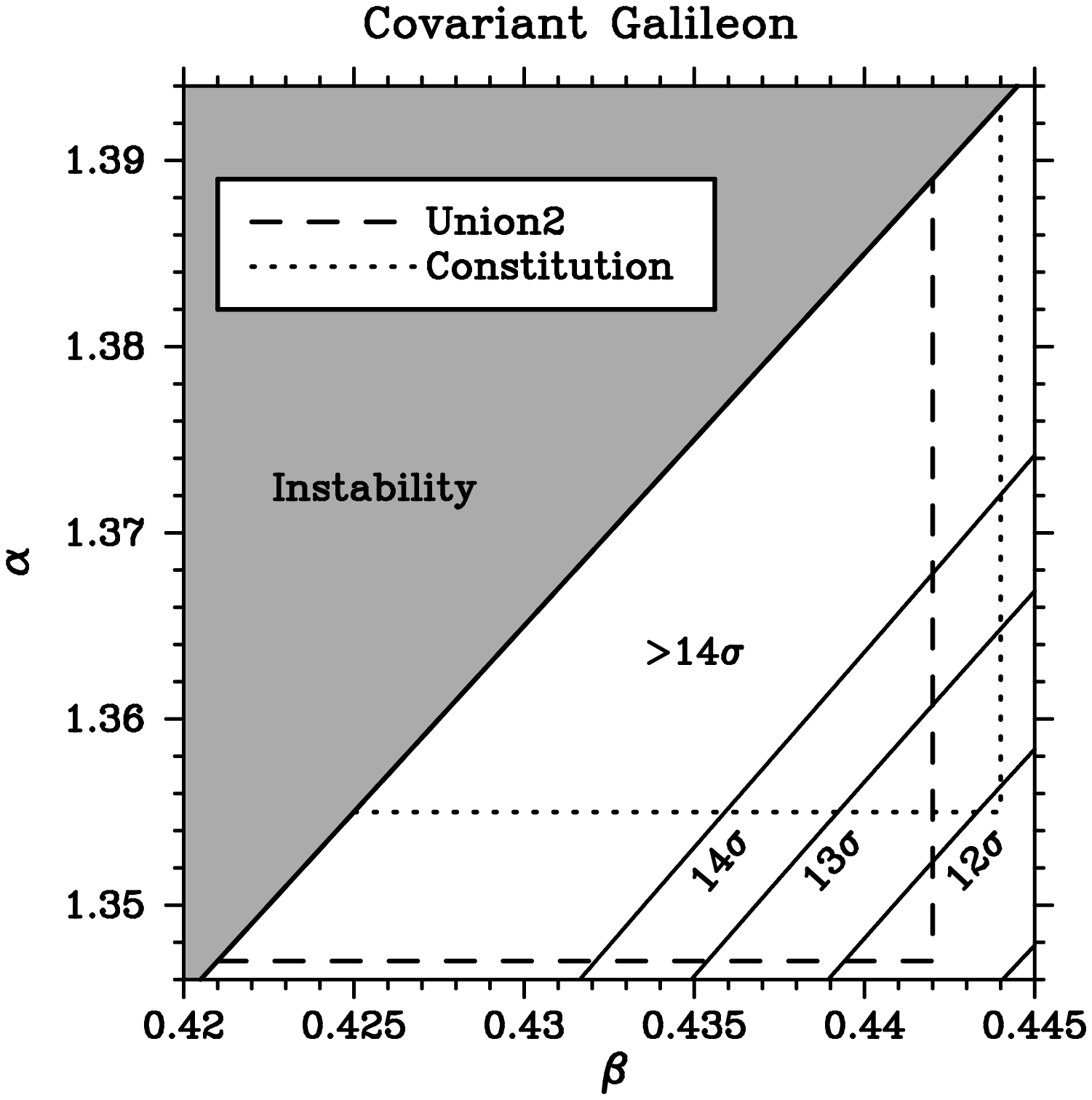} 
\par\end{centering}
\caption{Constraints on the covariant Galileon 
model given by the functions (\ref{GachoiceK})-(\ref{Gachoice}) from 
the RSD data in the $(\alpha,\beta)$ plane
for the scale $k^{-1} = 30\,h^{-1}$\,Mpc  
with $\sigma_8(z=0) 
= 0.811$ and $\Omega_{\rm DE}^{(0)} = 0.73$.
The dashed and dotted lines represent the regions
constrained by the background expansion history
according to the Union2 and Constitution SN Ia data, respectively.
In the grey region ($\alpha-2\beta>0.505$) there is an 
instability associated with negative values of $G_{\rm eff}$
during the transition from the matter era to the accelerated epoch. 
The parameter space that is theoretically valid and consistent with 
background-level observations is excluded by the RSD data
at more than 11 $\sigma$ CL.
}
\centering{} \label{fig7}
\end{figure}

For the covariant Galileon the allowed parameter space is constrained
as Eq.~(\ref{albe2}) from the combined data analysis of SN Ia (Union2),
CMB, and BAO.  If the Constitution SN Ia data are used
instead of Union2, the observational bounds on $\alpha$ and $\beta$ become
$\alpha=1.411 \pm 0.056$ and $\beta=0.422 \pm 0.022$ \cite{Nesseris}.
Setting the prior $\alpha-2\beta<0.505$ to avoid the negative values
of $G_{\rm eff}$, the allowed parameter space is the triangle
surrounded by the three points $(\alpha, \beta)=(1.347, 0.421),
(1.347, 0.442), (1.389, 0.442)$ for Union2 and $(\alpha,
\beta)=(1.355, 0.425), (1.355, 0.444), (1.393, 0.444)$ for
Constitution.  In these parameter regions we further place constraints
on the model from the RSD data.

In Fig.~\ref{fig7} we show observational bounds on the covariant 
Galileon model in the $(\alpha, \beta)$ plane for 
the scale $30\,h^{-1}$~Mpc with $\sigma_8(z=0) 
= 0.811$ and $\Omega_{\rm DE}^{(0)} = 0.73$.
The parameter space constrained from the background cosmology 
is excluded at more than $11 \sigma$ CL. 
This comes from the large cosmic growth rate relative to 
that of the $\Lambda$CDM, as shown in Fig.~\ref{fig3}.
As the model parameters approach the border 
$\alpha-2\beta=0.505$, it is further difficult to match with 
the data because of the significant enhancement 
of matter perturbations.
We also find that the observational constraints are 
practically independent of the scales.
For the scale $60\,h^{-1}$~Mpc the covariant Galileon 
is excluded at more than $10 \sigma$ CL.

A similar conclusion was recently reached in
Ref.~\cite{ApplebyLinder}, in which the authors included
the WiggleZ and BOSS data as well as SN Ia, CMB, BAO 
data. The difference from their work is that we have taken 
into account the 2dFGRS, 6dFGRS, and 
SDSS-LRG data as well. Even though the 2dFGRS and 
6dFGRS data provide larger values of $f_m \sigma_8$ 
relative to those from WiggleZ and BOSS, 
the covariant Galileon cannot be 
consistent with most of the currently available RSD data.

For smaller values of $\sigma_8(z=0)$, the constraint becomes looser.
In the range $0.75<\sigma_8(z=0)<0.85$, however, the whole parameter space
is still rejected at least at $8\sigma$ CL. 
If we take much smaller $\sigma_8 (z=0)$ such as $0.6$,
there could be some allowed region at $95.4\%$ CL.
However, such low values of $\sigma_8 (z=0)$ are disfavored from 
the constraint from clustering of galaxies \cite{Rapetti}.

\subsection{Extended Galileon}

In the extended Galileon model with $p=1$ and $q=5/2$, the
theoretically allowed parameter space is given by the region shown
in Fig.~\ref{fig4}.  As mentioned in the previous section, this plot
assumes that the dark energy equation of state during the matter era is
$w_{\rm DE}= -1 - s = -1.2$, where $s = p/(2q)$.  This model is
interesting because it is different from the $\Lambda$CDM model but
is still consistent with the joint constraint of SN Ia, CMB, 
and BAO \cite{DTconstraint} (see also Eq.~(\ref{eq:s_limit})).

\begin{figure}
\begin{centering}
\includegraphics[width=3.4in,height=3.4in]{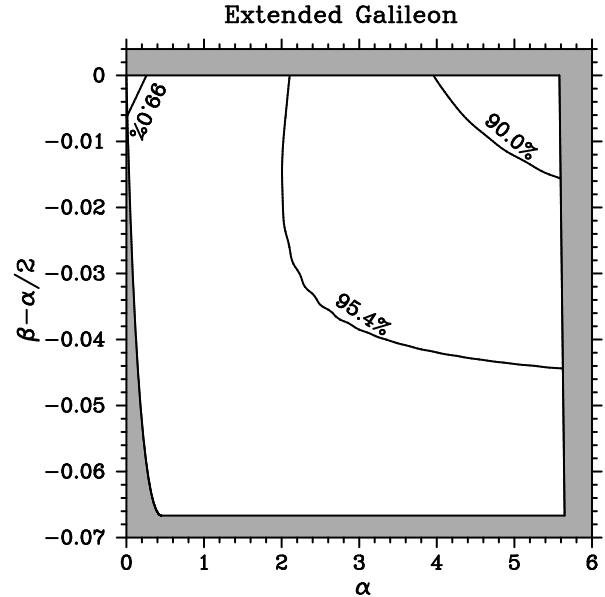} 
\par\end{centering}
\caption{Constraints on the extended Galileon 
model given by the functions (\ref{coex1})-(\ref{exGachoice}) from 
the RSD data in the $(\alpha,\beta)$ plane
for the scale $k^{-1} = 30\,h^{-1}$~Mpc
with $\sigma_8(z=0) 
= 0.811$ and $\Omega_{\rm DE}^{(0)} = 0.72$.
The grey scale region is theoretically excluded
(see Fig.~\ref{fig4}). 
The lower-left region separated by the 
95.4 \% contour is excluded at more than 95.4 \% CL, and
other contours have similar meanings. }
\centering{} \label{fig8}
\end{figure}

In Fig.~\ref{fig8} we plot the observational bounds in the 
$(\alpha, \beta)$ plane for the scale $30\,h^{-1}$~Mpc 
with $\sigma_8(z=0) 
= 0.811$ and $\Omega_{\rm DE}^{(0)} = 0.72$ 
(in which case the dark energy density parameter is consistent
with the background constraint derived in Ref.~\cite{DTconstraint}).
For the scales ranging 
$10\,h^{-1}$~Mpc $\lesssim\,k^{-1} \lesssim$ $60\,h^{-1}$~Mpc
we confirm that the constraints are 
similar to those shown in Fig.~\ref{fig8}.
Unlike the covariant Galileon there is a parameter space 
in which the model is within the $2\sigma$ observational
contour. 

{}From Fig.~\ref{fig8} we find that the models with larger 
values of $\alpha$ and $\beta$ 
tend to be favored due to the suppressed
growth of matter perturbations in high redshifts.
Case (iii) shown in Fig.~\ref{fig5} corresponds to 
such an example, in which case the model is 
inside the 90.0 \% CL observational contour. 
Case (i) in Fig.~\ref{fig5} is excluded at 
99.0 \% CL because of large values of $f_m \sigma_8$
in both high and low redshifts.
Case (ii) is outside the 95.4 \% CL contour
due to the large values of $f_m \sigma_8$ in 
high redshifts in spite of the suppressed evolution 
of $f_m \sigma_8$ for $z \lesssim 0.5$. 
In Table \ref{table:1Dconstraints} we summarize the results 
of one-dimensional constraints for a given value of 
$\alpha$ (or a combination $\beta-\alpha/2$).

Varying the dark energy density parameter in the range
$0.70<\Omega_{{\rm DE}}^{(0)}<0.75$ does not 
change the above constraints significantly.
Setting $\sigma_8(z=0) = 0.75$, we find that the whole 
parameter space can be allowed at $1\sigma$ CL. 
For lager $\sigma_8(z=0)$, observational constraints
get tighter.

It should be noted again that the observational bounds in
Fig.~\ref{fig8} correspond to the model parameter $s=0.2$.  For the
$s$ values close to $0$, the growth rate of matter perturbations gets
smaller and approaches the $\Lambda$CDM rate. Obviously such models
should also be compatible with the RSD data. Note that the
cross-correlation between the large-scale structure (LSS) and the
integrated-Sachs-Wolfe (ISW) effect in CMB can also provide additional
constraints on the extended Galileon model.  The joint data analysis
of the RSD and the LSS-ISW cross-correlation is beyond the scope of
our paper.

\section{Conclusions}
\label{consec}

In this paper we have placed observational constraints on dark energy
models based on $f(R)$ gravity, covariant Galileon, and extended
Galileon from the latest data of galaxy redshift surveys (WiggleZ,
SDSS LRG, BOSS, and 6dFGRS).  In these models the General Relativistic
behavior can be recovered at short distances under the chameleon
mechanism or the Vainshtein mechanism.  On scales relevant to
large-scale structures the modification of gravity manifests itself
for the growth rate of matter density perturbations. This growth rate
is related to the peculiar velocities of galaxies, which can be
constrained from the redshift-space distortions.

As an explicit example we studied the $f(R)$ model (\ref{fRmo})
proposed by Hu and Sawicki, which is general enough to cover basic
properties of $f(R)$ dark energy models.  At the level of
perturbations this model mimics the $\Lambda$CDM for the redshift $z$
larger than the critical value $z_c$ given by Eq.~(\ref{zc}), but the
deviation appears for $z<z_c$.  For smaller values of $n$ and
$\lambda$ the transition redshift $z_c$ gets larger, so that the
growth of matter perturbations is more significant at late times.

We placed observational bounds on the Hu-Sawicki $f(R)$ model in
the $(n, \lambda)$ plane.  Since the growth rate in this model
depends on the scale of density perturbations, we derive constraints
assuming three different wavenumbers of $k^{-1}=10, \ 30,$ and 
$60\, h^{-1}$~Mpc which are relevant to the galaxy redshift surveys
considered here.
As we see in Fig.~\ref{fig6} (which corresponds to 
$\sigma_8(z=0) = 0.811$ and $\Omega_{\rm DE}^{(0)} = 0.73$), 
the constraints on
model parameters tend to be tighter for smaller scales.  For
$k^{-1}=60 \, h^{-1}$~Mpc the models with $n<2$ and $\lambda<3$ are
outside the 2$\sigma$ observational contour, whereas for $k^{-1}=10 \,
h^{-1}$~Mpc the models with $n<2$ and $\lambda<8$ are excluded at the
2$\sigma$ CL.  These results show that the recent RSD data do not
favor the large deviation of $f_{m}\sigma_8$ from that in the
$\Lambda$CDM, whose property can be confirmed in Figs.~\ref{fig1} and
\ref{fig2}.

For covariant Galileon there is a tracker solution along which 
the equation of state $w_{\rm DE}$ of dark energy changes from 
$-2$ (matter era) to $-1$ (de Sitter era).
The likelihood analysis of SN Ia, CMB, and BAO 
shows that only the late-time tracking solution is allowed from the data.
Using the parameter space constrained from the background 
cosmology, we found that the covariant Galileon model is excluded
at more than $8 \sigma$ CL for $\sigma_8 (z=0)$ ranging 
in the region $0.75<\sigma_8 (z=0)<0.85$.
This is associated with the fact that the cosmic growth rate
for the covariant Galileon is much larger than that in the $\Lambda$CDM, 
as confirmed in Fig.~\ref{fig3}.
In the range $10\,h^{-1}$~Mpc $\lesssim\,k^{-1} \lesssim$ $60\,h^{-1}$~Mpc
the evolution of $f_m \sigma_8$ is practically independent of the scales.

In the extended Galileon scenario the equation of state for the tracker 
is given by $w_{\rm DE}=-1-s$ during the matter era, where $s$ is a positive 
constant. For the model with $s=0.2$ we found that there is 
a parameter space which is inside the $2\sigma$ 
observational contour. As we see in Fig.~\ref{fig8}, the models
with larger values of $\alpha$ and $\beta$ tend to be favored
from the RSD data.

According to the latest data release of Planck, the dark energy 
density parameter is constrained to be 
$\Omega_{\rm DE}^{(0)} = 0.686 \pm 0.020$ (68\,\%\,CL) \cite{Planck}.
Varying $\Omega_{\rm DE}^{(0)}$ in this range, the allowed regions
tend to be a little bit narrower in all three models.
However, this does not give rise to any qualitatively different change
to our results.

We have thus shown that the recent observations of redshift-space
distortions are already quite powerful to constrain a number of
modified gravity models.  In upcoming future we should obtain more $f
\sigma_8$ measurements that are more precise and/or at different
redshifts by larger data of ongoing surveys or near-future projects
such as Subaru/FMOS, Subaru/PFS, HETDEX and so on.  
We hope that we will be able to approach the origin of dark energy 
in the foreseeable future.

\section*{ACKNOWLEDGEMENTS}
The authors were supported by the Grant-in-Aid for Scientific Research
Fund of the JSPS (Grants No. 23684007 for TT and No. 24540286 for ST) and the
Fund for Scientific Research on Innovative Areas (JSPS Grant No.\,21111006
for ST).

\vspace{0.5cm}
{\it \underline{Note added.}}
\vspace{0.2cm}

After the first submission of this paper, several related papers appeared
\cite{Barreira,Li2013,Wyman,Neveu}.

\vspace{0.2cm}
 1. Barreira {\it et al.} \cite{Barreira} derived observational constraints on
the covariant Galileon model, but without using the RSD constraints.
They argued that the use of RSD constraints would induce significant
uncertainties by the non-linear effects of the Vainshtein mechanism.
Recently, Li {\it et al.} \cite{Li2013} showed that, in the DGP and Galileon models
with the field self-interaction $X \square \phi$, the non-linear effects start to 
affect the velocity field even for $k \sim 0.05$\,$h$\,Mpc$^{-1}$.
Wyman {\it et al.} \cite{Wyman} carried out the $N$-body simulations 
for a phenomenological model with a standard $\Lambda$CDM 
expansion history and a Galileon-type scalar field.
Although these analyses do not cover the models with the full 
covariant and extended Galileon terms discussed in this paper,  
it will be certainly of interest to study how the non-linear effects
can affect the constraints derived in our paper.

\vspace{0.2cm}
2. Neveu {\it et al.} \cite{Neveu} showed that, for the covariant Galileon, 
there is some allowed parameter space consistent with the RSD data.
They pointed out that this difference may come from
whether or not the Alcock-Paczynski effect is taken into account 
or from the difference of normalization of $\sigma_8 (z=0)$.
We confirmed that, unlike our work, Neveu {\it et al.} did not 
impose the existence of a future de Sitter solution.
In this case there is one free parameter ($c_2$ or $c_3$) 
in addition to $\alpha$ and $\beta$.
In fact, we checked that their best-fit model parameters do not 
belong to the theoretically allowed region constrained 
in Refs.~\cite{DTPRL,DTPRD}.
Our assumption for the existence of the de Sitter solution 
is reasonable, because the solutions with different initial conditions
converge to the attractor of cosmic acceleration.


\begin{thebibliography}{10}

\bibitem{SN98} 
A.~G.~Riess {\it et al.}  [Supernova Search Team Collaboration],
Astron.\ J.\  {\bf 116}, 1009 (1998);
S.~Perlmutter {\it et al.}  [Supernova Cosmology Project Collaboration],
Astrophys.\ J.\  {\bf 517}, 565 (1999).

\bibitem{CMB03}
D.~N.~Spergel {\it et al.}  [WMAP Collaboration],
Astrophys.\ J.\ Suppl.\  {\bf 148}, 175 (2003).

\bibitem{LSS04}
M.~Tegmark {\it et al.}  [SDSS Collaboration],
Phys.\ Rev.\  D {\bf 69}, 103501 (2004).

\bibitem{Sotiriou} 
T.~P.~Sotiriou and V.~Faraoni,
Rev.\ Mod.\ Phys.\  {\bf 82}, 451 (2010).

\bibitem{DT10} 
E.~J.~Copeland, M.~Sami and S.~Tsujikawa,
Int.\ J.\ Mod.\ Phys.\  D {\bf 15}, 1753 (2006);
S.~Tsujikawa,
Lect.\ Notes Phys.\  {\bf 800}, 99 (2010).

\bibitem{fRde} 
A.~De Felice and S.~Tsujikawa,
Living Rev.\ Rel.\  {\bf 13}, 3 (2010).

\bibitem{Clifton} 
T.~Clifton, P.~G.~Ferreira, A.~Padilla and C.~Skordis,
Phys.\ Rept.\  {\bf 513}, 1 (2012).

\bibitem{fR}
S.~Capozziello,
Int.\ J.\ Mod.\ Phys.\  D {\bf 11}, 483 (2002);
S.~Capozziello, S.~Carloni and A.~Troisi,
Recent Res.\ Dev.\ Astron.\ Astrophys.\  {\bf 1}, 625 (2003);
S.~M.~Carroll, V.~Duvvuri, M.~Trodden and M.~S.~Turner,
Phys.\ Rev.\  D {\bf 70}, 043528 (2004).

\bibitem{stensor}
L.~Amendola,
Phys.\ Rev.\  D {\bf 60}, 043501 (1999);
J.~P.~Uzan,
Phys.\ Rev.\  D {\bf 59}, 123510 (1999);
T.~Chiba,
Phys.\ Rev.\ D {\bf 60}, 083508 (1999);
N.~Bartolo and M.~Pietroni,
Phys.\ Rev.\ D {\bf 61} 023518 (1999);
F.~Perrotta, C.~Baccigalupi and S.~Matarrese,
Phys.\ Rev.\  D {\bf 61}, 023507 (1999).

\bibitem{DGP}
G.~R.~Dvali, G.~Gabadadze and M.~Porrati,
Phys.\ Lett.\  B {\bf 485}, 208 (2000).

\bibitem{Nicolis} 
A.~Nicolis, R.~Rattazzi and E.~Trincherini,
Phys.\ Rev.\ D {\bf 79}, 064036 (2009).

\bibitem{Deffayet} 
C.~Deffayet, G.~Esposito-Farese and A.~Vikman,
Phys.\ Rev.\ D {\bf 79}, 084003 (2009);
C.~Deffayet, S.~Deser and G.~Esposito-Farese,
Phys.\ Rev.\ D {\bf 80}, 064015 (2009).

\bibitem{fRper}
S.~M.~Carroll, I.~Sawicki, A.~Silvestri and M.~Trodden,
New J.\ Phys.\  \textbf{8}, 323 (2006);
R.~Bean, D.~Bernat, L.~Pogosian, A.~Silvestri and M.~Trodden,
Phys.\ Rev.\  D \textbf{75}, 064020 (2007);
L.~Pogosian and A.~Silvestri,
Phys.\ Rev.\ D {\bf 77}, 023503 (2008);
B.~Li and J.~D.~Barrow,
Phys.\ Rev.\ D {\bf 75}, 084010 (2007).

\bibitem{Tsujikawa07}
S.~Tsujikawa,
Phys.\ Rev.\  D {\bf 76}, 023514 (2007).

\bibitem{DGPper}
A.~Lue, R.~Scoccimarro and G.~D.~Starkman,
Phys.\ Rev.\  D {\bf 69}, 124015 (2004);
K.~Koyama and R.~Maartens,
JCAP {\bf 0601}, 016 (2006).

\bibitem{Kase} 
A.~De Felice, R.~Kase and S.~Tsujikawa,
Phys.\ Rev.\ D {\bf 83}, 043515 (2011).

\bibitem{DKT} 
A.~De Felice, T.~Kobayashi and S.~Tsujikawa,
Phys.\ Lett.\ B {\bf 706}, 123 (2011).

\bibitem{Horndeski} 
G.~W.~Horndeski, Int.\ J.\ Theor.\ Phys.\ 10, 363 (1974).

\bibitem{DGSZ} 
C.~Deffayet, X.~Gao, D.~A.~Steer and G.~Zahariade,
Phys.\ Rev.\  D {\bf 84}, 064039 (2011);
C.~Charmousis, E.~J.~Copeland, A.~Padilla and P.~M.~Saffin,
Phys.\ Rev.\ Lett.\  {\bf 108}, 051101 (2012).

\bibitem{KYY11} 
T.~Kobayashi, M.~Yamaguchi and J.~Yokoyama,
Prog.\ Theor.\ Phys.\  {\bf 126}, 511 (2011).

\bibitem{Kaiser} 
N.~Kaiser,
Mon.\ Not.\ Roy.\ Astron.\ Soc.\  {\bf 227}, 1 (1987).

\bibitem{Tegmark04} 
M.~Tegmark {\it et al.}  [SDSS Collaboration],
Astrophys.\ J.\  {\bf 606}, 702 (2004).

\bibitem{Percival04}
W.~J.~Percival {\it et al.}  [The 2dFGRS Collaboration],
Mon.\ Not.\ Roy.\ Astron.\ Soc.\  {\bf 353}, 1201 (2004).

\bibitem{Tegmark06} 
M.~Tegmark {\it et al.}  [SDSS Collaboration],
Phys.\ Rev.\ D {\bf 74}, 123507 (2006).

\bibitem{Yamamoto} 
K.~Yamamoto, T.~Sato and G.~Huetsi,
Prog.\ Theor.\ Phys.\  {\bf 120}, 609 (2008).

\bibitem{Guzzo08} 
L.~Guzzo  {\it et al.},
Nature {\bf 451}, 541 (2008).

\bibitem{Blake} 
C.~Blake {\it et al.},
Mon.\ Not.\ Roy.\ Astron.\ Soc.\  {\bf 415}, 2876 (2011).

\bibitem{Samushia11} 
L.~Samushia, W.~J.~Percival and A.~Raccanelli,
Mon.\ Not.\ Roy.\ Astron.\ Soc.\  {\bf 420}, 2102 (2012).

\bibitem{Reid12} 
B.~A.~Reid, L.~Samushia, M.~White, W.~J.~Percival, M.~Manera, 
N.~Padmanabhan, A.~J.~Ross and A.~G.~Sanchez {\it et al.},
Mon.\ Not.\ Roy.\ Astron.\ Soc.\  {\bf 426}, 2719 (2012).

\bibitem{Beutler12} 
F.~Beutler, C.~Blake, M.~Colless, D.~H.~Jones, L.~Staveley-Smith, G.~B.~Poole, 
L.~Campbell and Q.~Parker {\it et al.},
arXiv:1204.4725 [astro-ph.CO].

\bibitem{Samushia12} 
L.~Samushia, B.~A.~Reid, M.~White, W.~J.~Percival, A.~J.~Cuesta, 
L.~Lombriser, M.~Manera and R.~C.~Nichol {\it et al.},
Mon.\ Not.\ Roy.\ Astron.\ Soc.\  {\bf 429}, 1514 (2013).

\bibitem{Hudson} 
M.~J.~Hudson and S.~J.~Turnbull,
Astrophys.\ J.\  {\bf 751}, L30 (2012).

\bibitem{Jailson} 
S.~Tsujikawa, A.~De Felice and J.~Alcaniz,
JCAP {\bf 1301}, 030 (2013).

\bibitem{chame}
J.~Khoury and A.~Weltman,
Phys.\ Rev.\ Lett.\  {\bf 93}, 171104 (2004);
Phys.\ Rev.\  D {\bf 69}, 044026 (2004).

\bibitem{Vainshtein}
A.~I.~Vainshtein,
Phys.\ Lett.\  B {\bf 39}, 393 (1972).

\bibitem{symmetrons}
K.~Hinterbichler and J.~Khoury,
Phys.\ Rev.\ Lett.\  {\bf 104}, 231301 (2010);
K.~Hinterbichler, J.~Khoury, A.~Levy and A.~Matas,
Phys.\ Rev.\ D {\bf 84}, 103521 (2011).

\bibitem{BBDS}
P.~Brax, C.~van de Bruck, A.~-C.~Davis and D.~Shaw,
Phys.\ Rev.\ D {\bf 82}, 063519 (2010).

\bibitem{Brachame}
S.~Tsujikawa, K.~Uddin, S.~Mizuno, R.~Tavakol and J.~Yokoyama,
Phys.\ Rev.\ D {\bf 77}, 103009 (2008);
R.~Gannouji, B.~Moraes, D.~F.~Mota, D.~Polarski, S.~Tsujikawa and H.~A.~Winther,
Phys.\ Rev.\ D {\bf 82}, 124006 (2010).

\bibitem{Song07} 
Y.~-S.~Song, W.~Hu and I.~Sawicki,
Phys.\ Rev.\ D {\bf 75}, 044004 (2007).

\bibitem{Peiris} 
Y.~-S.~Song, H.~Peiris and W.~Hu,
Phys.\ Rev.\ D {\bf 76}, 063517 (2007).

\bibitem{Lombriser} 
L.~Lombriser, A.~Slosar, U.~Seljak and W.~Hu,
Phys.\ Rev.\ D {\bf 85}, 124038 (2012).

\bibitem{fRobsercon} 
K.~Koyama, A.~Taruya and T.~Hiramatsu,
Phys.\ Rev.\ D {\bf 79}, 123512 (2009);
G.~B.~Zhao {\it et al.},
Phys.\ Rev.\ D {\bf 81}, 103510 (2010);
T.~Narikawa and K.~Yamamoto,
Phys.\ Rev.\ D {\bf 81}, 043528 (2010);
K.~Yamamoto, G.~Nakamura, G.~Hutsi, T.~Narikawa and T.~Sato,
Phys.\ Rev.\ D {\bf 81}, 103517 (2010).

\bibitem{Jennings} 
E.~Jennings, C.~M.~Baugh, B.~Li, G.~-B.~Zhao and K.~Koyama,
Mon.\ Not.\ Roy.\ Astron.\ Soc.\  {\bf 425}, 2128 (2012);
B.~Li, W.~A.~Hellwing, K.~Koyama, G.~-B.~Zhao, E.~Jennings and C.~M.~Baugh,
Mon.\ Not.\ Roy.\ Astron.\ Soc.\  {\bf 428}, 743 (2013).

\bibitem{GS10} 
R.~Gannouji and M.~Sami,
Phys.\ Rev.\ D {\bf 82}, 024011 (2010).

\bibitem{DTPRL} 
A.~De Felice and S.~Tsujikawa,
Phys.\ Rev.\ Lett.\  {\bf 105}, 111301 (2010).

\bibitem{DTPRD} 
A.~De Felice and S.~Tsujikawa,
Phys.\ Rev.\ D {\bf 84}, 124029 (2011).

\bibitem{ApplebyLinder} 
S.~A.~Appleby and E.~V.~Linder,
JCAP {\bf 1208}, 026 (2012).

\bibitem{Will} 
C.~M.~Will,
Living Rev.\ Rel.\  {\bf 9}, 3 (2005).

\bibitem{Star80}
A.~A.~Starobinsky,
Phys.\ Lett.\ B {\bf 91}, 99 (1980).

\bibitem{fRchame}
J.~A.~R.~Cembranos,
Phys.\ Rev.\  D {\bf 73}, 064029 (2006);
I.~Navarro and K.~Van Acoleyen,
JCAP {\bf 0702}, 022 (2007);
T.~Faulkner, M.~Tegmark, E.~F.~Bunn and Y.~Mao,
Phys.\ Rev.\  D {\bf 76}, 063505 (2007).

\bibitem{Capo} 
S.~Capozziello and S.~Tsujikawa,
Phys.\ Rev.\ D {\bf 77}, 107501 (2008).

\bibitem{HuSa} 
W.~Hu and I.~Sawicki,
Phys.\ Rev.\ D {\bf 76}, 064004 (2007).

\bibitem{Star07} 
A.~A.~Starobinsky,
JETP Lett.\  {\bf 86}, 157 (2007).

\bibitem{Appleby} 
S.~A.~Appleby and R.~A.~Battye,
Phys.\ Lett.\ B {\bf 654}, 7 (2007).

\bibitem{Tsuji07} 
S.~Tsujikawa,
Phys.\ Rev.\ D {\bf 77}, 023507 (2008).

\bibitem{Linder} 
E.~V.~Linder,
Phys.\ Rev.\ D {\bf 80}, 123528 (2009).

\bibitem{DGPVain1}
C.~Deffayet, G.~R.~Dvali, G.~Gabadadze and A.~I.~Vainshtein,
Phys.\ Rev.\  D {\bf 65}, 044026 (2002).

\bibitem{DGPVain2}
M.~Porrati,
Phys.\ Lett.\  B {\bf 534}, 209 (2002);
M.~A.~Luty, M.~Porrati and R.~Rattazzi,
JHEP {\bf 0309}, 029 (2003).

\bibitem{DGPghost}
A.~Nicolis and R.~Rattazzi,
JHEP {\bf 0406}, 059 (2004);
D.~Gorbunov, K.~Koyama and S.~Sibiryakov,
Phys.\ Rev.\  D {\bf 73}, 044016 (2006).

\bibitem{DGPcon}
M.~Fairbairn and A.~Goobar,
Phys.\ Lett.\  B {\bf 642}, 432 (2006);
R.~Maartens and E.~Majerotto,
Phys.\ Rev.\  D {\bf 74}, 023004 (2006);
U.~Alam and V.~Sahni,
Phys.\ Rev.\  D {\bf 73}, 084024 (2006);
Y.~S.~Song, I.~Sawicki and W.~Hu,
Phys.\ Rev.\  D {\bf 75}, 064003 (2007);
J.~Q.~Xia,
Phys.\ Rev.\  D {\bf 79}, 103527 (2009).

\bibitem{Nesseris} 
S.~Nesseris, A.~De Felice and S.~Tsujikawa,
Phys.\ Rev.\ D {\bf 82}, 124054 (2010).

\bibitem{Kimura} 
R.~Kimura and K.~Yamamoto,
JCAP {\bf 1104}, 025 (2011).

\bibitem{DTmodel} 
G.~Dvali and M.~S.~Turner,
arXiv:astro-ph/0301510.

\bibitem{KKY} 
R.~Kimura, T.~Kobayashi and K.~Yamamoto,
Phys.\ Rev.\ D {\bf 85}, 123503 (2012).

\bibitem{DTcondition} 
A.~De Felice and S.~Tsujikawa,
JCAP {\bf 1202}, 007 (2012).

\bibitem{DTconstraint} 
A.~De Felice and S.~Tsujikawa,
JCAP {\bf 1203}, 025 (2012).

\bibitem{fRearly}
T.~V.~Ruzmaikina and A.~A.~Ruzmaikin, Zh.\ Eksp.\ Teor.\ Fiz. {\bf 57},
680 (1969) [Sov. Phys. - JETP {\bf 30}, 372 (1970)]; 
B.~N.~Breizman, V.~Ts.~Gurovich and V.~P.~Sokolov, 
Zh.\ Eksp.\ Teor.\ Fiz. {\bf 59}, 288 (1970) 
[Sov. Phys. - JETP {\bf 32}, 155 (1971)];
R.~Kerner,
Gen.\ Rel.\ Grav.\  {\bf 14}, 453 (1982).

\bibitem{Muller} 
V.~Muller, H.~J.~Schmidt and A.~A.~Starobinsky,
Phys.\ Lett.\ B {\bf 202}, 198 (1988).

\bibitem{Faraoni} 
V.~Faraoni,
Phys.\ Rev.\ D {\bf 72}, 061501 (2005);
V.~Faraoni and S.~Nadeau,
Phys.\ Rev.\ D {\bf 72}, 124005 (2005).

\bibitem{Amendola} 
L.~Amendola, R.~Gannouji, D.~Polarski and S.~Tsujikawa,
Phys.\ Rev.\ D {\bf 75}, 083504 (2007);
L.~Amendola and S.~Tsujikawa,
Phys.\ Lett.\ B {\bf 660}, 125 (2008).

\bibitem{Barrow} 
B.~Li and J.~D.~Barrow,
Phys.\ Rev.\ D {\bf 75}, 084010 (2007).

\bibitem{Bardeen} 
J.~M.~Bardeen,
Phys.\ Rev.\ D {\bf 22}, 1882 (1980).

\bibitem{Star98} 
A.~A.~Starobinsky, 
JETP Lett.\ \textbf{68}, 757 (1998); 
B.~Boisseau, G.~Esposito-Farese, D.~Polarski and A.~A.~Starobinsky, 
Phys.\ Rev.\ Lett.\ \textbf{85}, 2236 (2000); 
A.~De Felice, S.~Mukohyama and S.~Tsujikawa, 
Phys.\ Rev.\ \textbf{D82}, 023524 (2010).

%

\bibitem{Komatsu} 
E.~Komatsu {\it et al.}  [WMAP Collaboration],
Astrophys.\ J.\ Suppl.\  {\bf 192}, 18 (2011).

\bibitem{Rapetti}
D.~Rapetti, C.~Blake, S.~W.~Allen, A.~Mantz, D.~Parkinson and F.~Beutler,
arXiv:1205.4679 [astro-ph.CO].

\bibitem{TGMP} 
S.~Tsujikawa, R.~Gannouji, B.~Moraes and D.~Polarski,
Phys.\ Rev.\ D {\bf 80}, 084044 (2009).

\bibitem{Tavakol} 
S.~Tsujikawa, K.~Uddin and R.~Tavakol,
Phys.\ Rev.\ D {\bf 77}, 043007 (2008).

\bibitem{Motohashi} 
H.~Motohashi, A.~A.~Starobinsky and J.~Yokoyama,
Int.\ J.\ Mod.\ Phys.\ D {\bf 18}, 1731 (2009).

\bibitem{ApplebyJCAP} 
S.~Appleby and E.~V.~Linder,
JCAP {\bf 1203}, 043 (2012).

\bibitem{Barreira12} 
A.~Barreira, B.~Li, C.~M.~Baugh and S.~Pascoli,
Phys.\ Rev.\ D {\bf 86}, 124016 (2012).

\bibitem{Barreira} 
A.~Barreira, B.~Li, A.~Sanchez, C.~M.~Baugh and S.~Pascoli,
arXiv:1302.6241 [astro-ph.CO].

\bibitem{Planck}
P. A. R. Ade {\it et al.}  [Planck Collaboration],
arXiv:1303.5076 [astro-ph.CO].

\bibitem{Li2013} 
B.~Li, G.~-B.~Zhao and K.~Koyama,
arXiv:1303.0008 [astro-ph.CO].

\bibitem{Wyman}
M.~Wyman, E.~Jennings and M.~Lima,
arXiv:1303.6630 [astro-ph.CO].

\bibitem{Neveu}
J.~Neveu, V.~Ruhlmann-Kleinder, A.~Conley, 
N.~Palanque-Delabrouille, P.~Asitier, J.~Guy and E.~Babichev, 
arXiv:1302.2786 [gr-qc].

\end{thebibliography}
\end{document}